\documentclass[12pt,preprint]{aastex}

\begin{document}

\title{Molecular Hydrogen in the Damped Ly$\alpha$ Absorber of Q1331+170}
\author{Jun Cui\altaffilmark{1}, Jill Bechtold\altaffilmark{1}, Jian Ge\altaffilmark{2},
and David M. Meyer\altaffilmark{3}}
\altaffiltext{1}{Steward Observatory, University of Arizona,
933 N Cherry Avenue, Tucson, AZ 85721, USA; Email: jcui@as.arizona.edu}
\altaffiltext{2}{Dept. of Astronomy, University of Florida,
210 Bryant Space Science Center, Gainesville, FL 32611, USA}
\altaffiltext{3}{Dept. of Physics and Astronomy, Northwestern University,
2145 Sheridan Road, Evanston, IL 60208, USA}

\begin{abstract}

We used HST/STIS to obtain the spectrum of molecular hydrogen associated
with the damped Ly$\alpha$ system at $z_{\rm abs}=1.7765$ toward the quasar Q1331+170
at $z_{\rm em}=2.084$. Strong ${\rm H}_2$ absorption was detected, with a total ${\rm H}_2$
column density of $N({\rm H}_2)=(4.45\pm 0.36)\times 10^{19}\:{\rm cm^{-2}}$.
The molecular hydrogen fraction is
$f_{{\rm H}_2}=\frac{2N_{\rm H_2}}{N_{\rm HI}+2N_{\rm H_2}}=(5.6\pm 0.7)\%$, which
is the greatest value reported so far in any redshifted damped Ly$\alpha$
system. This results from the combined effect of a relatively high dust-to-gas ratio,
a low gas temperature, and an extremely low ambient UV radiation field.
Based on the observed population of $J$ states,
we estimate the photo-absorption rate to be $R_{\rm abs}=(7.6\pm 2.4)\times 10^{-13}\:{\rm s^{-1}}$,
corresponding to a local UV radiation field of
$J(1000{\rm \AA})\approx 2.1\times 10^{-3} J_{1000{\rm \AA},\odot}$,
where $J_{1000{\rm \AA},\odot}$ is the UV intensity at $1000\:\AA$ in the solar neighborhood.
This is comparable with the metagalactic UV background intensity at this redshift,
and implies an extremely low star formation rate in the absorber's environment.
We construct a simple model to describe the structure of the ${\rm H}_2$ absorber,
with a best-fit total hydrogen number density of $n({\rm H})\approx 0.2\:{\rm cm^{-3}}$ and an electron
temperature of $T_{\rm e}\approx 140\:{\rm K}$. Assuming spherical symmetry,
the mass of the ${\rm H}_2$ cloud is estimated to be $\approx 6.5\times 10^7\:{\rm M_\odot}$,
larger than the masses of most giant molecular clouds (GMCs) in the Milky Way and nearby galaxies.
The extinction of Q1331+170 due to the intervening DLA is $E_{\rm B-V}\approx 0.037$, and we
also find that the extinction by DLAs with firm ${\rm H}_2$ detections is significantly greater
than those for which only upper limits of $f_{\rm H_2}$ have been put.
The observed CO-to-H$_2$ column density ratio is $\frac{N_{\rm CO}}{N_{\rm H_2}}<2.5\times 10^{-7}$,
which is similar to the value measured for diffuse molecular clouds in the Galactic ISM.
Finally, applying the inferred
physical conditions to the observed C~I fine structure excitation (Songaila {\it et al.} 1994),
we estimate the cosmic microwave background temperature to be
$T_{\rm CMB}=(7.2\pm 0.8)\:{\rm K}$ at $z=1.77654$, consistent with the
predicted value of $7.566\:{\rm K}$ from the standard cosmology.

\end{abstract}


\section{Introduction}

Damped Ly$\alpha$ absorbers (hereafter DLAs) are defined as quasar absorption systems with neutral hydrogen
column density, $N({\rm HI})>2\times 10^{20}\:{\rm cm^{-2}}$.
They are generally considered to be progenitors of present-day galaxies, in the form
of fast rotating disks (Prochaska \& Wolfe 1997), or merging protogalactic clumps (Haehnelt
{\it et al.} 1998). They serve as an important gas reservoir for star formation
at high redshifts (e.g. Storrie-Lombardi \& Wolfe 2000).

Molecular gas, especially molecular hydrogen, is an important ingredient of
star formation, since stars form by cooling via molecular lines.
The search for ${\rm H}_2$ in DLAs can be carried out by observing the ${\rm H}_2$ absorption lines 
in the Lyman X$^1\Sigma^+_g\longrightarrow$B$^1\Sigma^+_u$ and Werner
X$^1\Sigma^+_g\longrightarrow$C$^1\Sigma^{\pm}_u$ bands,
with rest wavelengths from $\sim$ 1220\AA\ to the Lyman limit (Abgrall {\it et al.} 1993a, 1993b).
Ge \& Bechtold (1999) conducted a survey for ${\rm H}_2$ absorption
in 13 DLAs, using the MMT Spectrograph. Two have been
confirmed with significant ${\rm H}_2$ fraction;
upper limits in the range of $10^{-6}$ to $10^{-4}$ were put
on the ${\rm H}_2$ fraction in other systems (Ge \& Bechtold 1997, 1999, Ge {\it et al.} 2001).
More recently, based on the high resolution
spectra obtained with UVES at the ESO VLT, Ledoux {\it et al.} (2003) searched
for ${\rm H}_2$ in a sample of 33 DLAs, with firm detections of ${\rm H}_2$ absorption
in 8 of them. By contrast, almost every line of sight through Galactic or Magellanic
gas clouds with comparable H~I column density has
strong ${\rm H}_2$ absorption (e.g. Shull {\it et al.} 2000).

In the interstellar medium, ${\rm H}_2$ molecules are expected to form
on the surface of dust grains by physical or chemical adsorption (Cazaux \& Tielens 2002).
Therefore one contributing factor to the lack of ${\rm H}_2$ in DLAs is the low
dust-to-gas ratios in these systems (e.g. Pei {\it et al.} 1991, Pettini {\it et al.} 1994, Vladilo 1998,
Prochaska \& Wolfe 2002), perhaps the result of the
fact that quasar surveys tend to select objects which are bright and blue (e.g. Gregg {\it et al.} 2002, Khare {\it et al.} 2004).
A second contributing factor to the lack of ${\rm H}_2$ may be the
high gas temperature in DLAs which makes ${\rm H}_2$ formation onto dust
grains inefficient (Petitjean {\it et al.} 2000,
Liszt 2002). This is consistent with the suggestion that unlike the local ISM, the gas
in DLAs might be predominantly warm
(Lu {\it et al.} 1996, Prochaska \& Wolfe 1999). Finally, ${\rm H}_2$ molecules are easily dissociated by UV
photons in the energy range of 11.3 --- 13.6 eV. Therefore the low ${\rm H}_2$ fraction in DLAs may be the result of
a strong UV radiation field (e.g. Black {\it et al.} 1987),
which could in turn be a signature of high star formation rate (Wolfe {\it et al.} 2003a, 2003b, 2004).
Recently, Hirashita {\it et al.} (2003) showed that the deficiency of ${\rm H}_2$ in DLAs may also
be a result of the expected low volume filling factor of ${\rm H}_2$ in environments with a strong UV radiation
field and low dust-to-gas ratio, rather than a real absence of ${\rm H}_2$.

Q1331+170 (at $z_{\rm em}=2.084$) is one of the first high redshift quasars discovered
(Baldwin {\it et al.} 1973), and has a damped Ly$\alpha$ absorber at $z=1.7765$
(Carswell {\it et al.} 1975) with a neutral hydrogen column density
of $N_{\rm HI}=1.5\times 10^{21}\:{\rm cm^{-2}}$ (Chaffee {\it et al.} 1988, Prochaska \& Wolfe 1999).
The 21-cm absorption line was also detected
(Wolfe \& Davis 1979, Chengalur \& Kanekar 2000), with a spin temperature of $T_{\rm s}\approx 770\:{\rm K}$.
This implies the presence of a cold neutral medium (CNM) phase
conducive to the efficient formation of ${\rm H}_2$ onto dust grains (Petitjean {\it et al.} 2000).
One of the most interesting features in the absorption spectrum of Q1331+170 is the presence of
strong C~I and C~I* multiple lines near $\lambda\sim 1656\:{\rm \AA}$
(Meyer {\it et al.} 1986, Songaila {\it et al.} 1994). 
C~I absorption has been found to be rare in quasar absorbers as compared to Galactic
diffuse clouds as a result of higher UV radiation fields, or lower dust abundances
(Ge {\it et al.} 1997, 2001). Since these conditions would also be inhospitable to an appreciable
formation of ${\rm H}_2$, the damped Ly$\alpha$ absorber toward Q1331+170
is an excellent candidate for a search for ${\rm H}_2$ absorption. 

Detecting C~I is also of cosmological interest, since it is insensitive to
the local physical conditions, but depends strongly on
the cosmic background level (Liszt 2002). Therefore the fine structure transitions
of C~I can be used to measure the CMB temperature as a function of redshift
(e.g. Bahcall {\it et al.} 1973). In the case of the DLA toward Q1331+170,
the excitation temperature of neutral carbon determined by Songaila
{\it et al.} (1994) is $7.4\pm 0.8\:{\rm K}$ for one of two velocity components at $z=1.77654$,
in agreement with the predicted value of $T_{\rm CMB}=7.566\:{\rm K}$ at that redshift.
This result implies that local excitation of C~I is negligible.
Since the population of ${\rm H}_2$ at different $J$ states can also be used to determine physical
parameters, such as kinetic temperature, neutral hydrogen number density, and UV radiation field,
observations of molecular hydrogen and its rotational excitation
provide an independent verification of the observed C~I excitation.

In this paper, we present our detection of redshifted ${\rm H}_2$ in the high resolution
absorption spectrum of Q1331+170 taken with the {\it Space Telescope Imaging
Spectrograph} (STIS) aboard the {\it Hubble Space Telescope} (HST).
Sec. 2 describes the observations and basic data reductions. In Sec. 3,
we describe the Voigt profile
fitting of ${\rm H}_2$ lines, and present the best-fit physical parameters.
Discussions are presented in Sec. 4, and conclusions in Sec. 5.

\section{Observations and continuum fitting}

In HST GO programs 7271 and 9172, we used HST/STIS to obtain the near UV spectrum of Q1331+170
in the wavelength range of 2500\AA\ to 3120\AA. The spectrum was obtained with
the E230M grating with $0.2^{\arcsec}\times 0.2^{\arcsec}$ aperture, and the NUV/MAMA detector
in the ACCUM operating mode, resulting in spectral resolution of $R=30,000$
(or $10\:{\rm km\:s^{-1}}$) at $\lambda\sim$ 2700\AA.
The total exposure time was 52,298 seconds, yielding spectra with S/N$\sim$7 above the Lyman
limit of the damped Ly$\alpha$ absorber ($\sim$2530\AA). The
data sets are listed in Table 1.

\begin{deluxetable}{cccc}
\tablecaption{Log of HST/STIS observations\label{tb1}}
\tablewidth{0pt}
\tablehead{
\colhead{Dataset}&\colhead{Program}&\colhead{UT date of observation}&\colhead{Exposure (sec)}}
\startdata
o56m0101&7271&1999-08-20&2076\\
o56m0102&7271&1999-08-20&5794\\
o56m0201&7271&1999-08-21&2076\\
o56m0202&7271&1999-08-21&5794\\
o6cy0101&9172&2002-07-08&2097\\
o6cy0102&9172&2002-07-08&5794\\
o6cy0201&9172&2002-08-12&2097\\
o6cy0202&9172&2002-08-12&5794\\
o6cy0301&9172&2002-08-19&2097\\
o6cy0302&9172&2002-08-19&5794\\
o6cy0401&9172&2002-08-19&2097\\
o6cy0402&9172&2002-08-19&5794\\
o6cy0501&9172&2003-03-16&2097\\
o6cy0502&9172&2003-03-16&5794\\
o6cy0601&9172&2003-03-12&2097\\
o6cy0602&9172&2003-03-12&5794\\
o6cy0701&9172&2002-07-10&2097\\
o6cy0702&9172&2002-07-10&2897\\
\enddata
\end{deluxetable}

The data were pre-processed with the IRAF calibration routine CALSTIS (v2.7) for
nonlinearity correction, dark subtraction, flat-fielding, wavelength
calibration, and extraction of the 1-D spectrum. Rebinning was performed on the raw
images for Nyquist sampling (i.e. 2 pixels per resolution element).
Instead of using automatic wavelength calibration files taken for each
Mode Selection Mechanism (MSM) setting, we obtained additional exposures
of comparison lamps during each orbit.
Typical observing conditions may cause thermal drifts of $0.1 \:{\rm pixel}$
per hour (see Sec. 11.2 of {\it Space Telescope Imaging Spectrograph Instrument Handbook
for Cycle 13} for more details).
For our longest single exposure, this corresponds to a wavelength uncertainty
of $\sim 0.007 {\rm \AA}$ at $\lambda\sim 2700{\rm \AA}$ introduced by wavelength calibration,
or a velocity uncertainty of $\sim 0.8\:{\rm km\:s^{-1}}$.

\begin{figure*}
\figurenum{1}
\epsscale{1.1}
\plotone{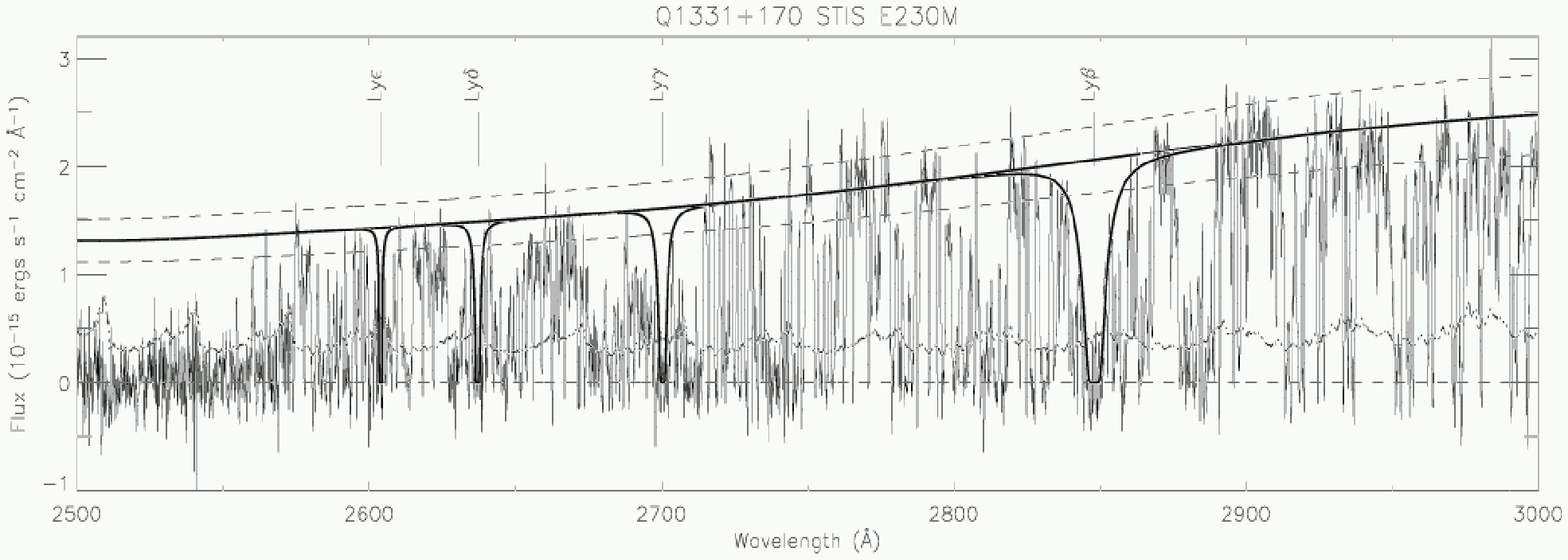}
\caption{The STIS E230M spectrum of Q1331+170.
The best-fit continuum (the thick solid line) is superposed on the flux-calibrated
spectrum. The dotted line corresponds to $1\:\sigma$ uncertainties
of the observed spectrum. The data have been smoothed by 5 pixels.
The two dashed lines represent the best-fit continuum scaled $15\%$ lower and
$15\%$ higher.
Also shown on the figure are the synthetic profiles of Lyman series lines,
including Ly$\beta$, Ly$\gamma$, Ly$\delta$ and Ly$\epsilon$. The synthetic profiles
are calculated with a neutral hydrogen column density of $1.5\times 10^{21}\:{\rm cm^{-2}}$
and a $b$-parameter of $19.6\:{\rm km\:s^{-1}}$ (see text).}
\end{figure*}

The entire flux-calibrated spectrum of Q1331+170 is shown in Fig. 1, smoothed by 5 pixels.
The continuum was determined by a 4th order polynomial fit using the data longward of the
Lyman limit of the DLA ($\approx$ 2530\AA). Regions with strong absorption features
identified by eye were excluded.
The dotted line in Fig. 1 shows the $1\:\sigma$ uncertainties of the observed spectrum.
The continuum fit suffers from
considerable uncertainties, due to the numerous absorption features
in the Ly$\alpha$ forest, as well as ${\rm H}_2$ absorption lines in the Lyman and Werner bands.
For this reason, we checked our continuum fit according to the following prescription.
We separate the whole spectrum into several continuous wavelength regions,
and for each region, we construct the histogram of all the data points with normalized flux
greater than 1. Ideally, the histogram can be well
represented by a (half-)Gaussian distribution with a standard deviation corresponding to the
normalized $1\:\sigma$ error of our spectrum. Next, we adjust the best-fit continuum in the region
under consideration until the calculated Gaussian standard deviation is approximately equal to the average
normalized $1\:\sigma$ error in the same region.
The above procedure was performed in an iterative manner, since an adjustment of the
continuum also changes the (normalized) $1\:\sigma$ error used for comparison.
This procedure can be used to determine both the overall amplitude and shape of the continuum.

We estimate the accuracy of continuum fitting by calculating the reduced
$\chi^2$-deviation between the probability distribution function (PDF) of normalized
flux adopting the best-fit continuum and that with continuum level scaled by a certain factor
(ranging from 50\% lower to 50\% higher). Here only pixels with normalized flux greater
than 1 are considered. This calculation shows that the best-fit continuum is accurate within
15\% at 83\% confidence level, and accurate within 25\% at 98\% confidence level,
based on the $\chi^2$ distribution.

The two dashed lines in Fig. 1 correspond to the best-fit continuum scaled 15\% lower and
15\% higher. This represents a very conservative estimate of the uncertainty in the continuum fit;
clearly the dashed lines are inconsistent with the expected continuum level in portions of the spectrum.
The effects of the possible uncertainties in continuum fitting on our final
results are discussed in Sec. 3.3.

\section{Analysis of molecular hydrogen lines}

\subsection{Detection of molecular hydrogen}

Previous spectroscopic studies show that there are at
least five redshift systems with detectable metal absorption lines along the line of sight
toward Q1331+170. They include the damped Ly$\alpha$ system at $z=1.7765$,
as well as absorbers at $z=0.7441$ (Sargent {\it et al.} 1988),
$z=1.3284$ (Steidel \& Sargent 1992), $z=1.4458$ (Sargent {\it et al.} 1988)
and $z=1.7864$ (Lanzetta {\it et al.} 1987).

We searched for molecular hydrogen absorption associated with the
$z=1.7765$ DLA and two other absorbers at $z=1.4458$ and $z=1.7864$, since their redshifts
allow considerable overlap between the wavelength coverage of our spectrum and the redshifted
Lyman and Werner bands of ${\rm H}_2$.
By comparing with the synthetic ${\rm H}_2$ spectrum, we detected strong molecular hydrogen
absorption associated with the $z=1.7765$ damped Ly$\alpha$ system, whereas the ${\rm H}_2$
line pattern does not match the observed spectrum for the other two absorbers.

For the damped Ly$\alpha$ system at $z=1.7765$, we calculated the
significance level of ${\rm H}_2$ lines for each $J$ state. We define the
significance level, ${\rm SL}$ associated with state $J$ as
\begin{equation}
{\rm SL}=\frac{\sum_i\int[1-f_i(\lambda)]d\lambda}{[\sum_i\int\sigma_i^2(\lambda)d\lambda]^{1/2}},
\end{equation}
\noindent
where $f_i(\lambda)$ and $\sigma_i(\lambda)$ are the normalized flux and $1\:\sigma$
error within an individual ${\rm H}_2$ line, and
the summation is over all the selected ${\rm H}_2$ lines for the same $J$ state.
The results show that based on our spectrum,
$J=0, 1, 2, 3, 4$ lines are detectable at $4.4\:\sigma$,
$5.1\:\sigma$, $9.0\:\sigma$, $8.9\:\sigma$ and $3.7\:\sigma$ significance level,
respectively. The $J\ge 5$ lines were not detected at $3\:\sigma$ significance level.
The number of unblended ${\rm H}_2$ lines included in the above calculation is
two for $J=0$, three for $J=1$, six for $J=2$, ten for $J=3$, two for $J=4$, and five for $J=5$, respectively
(see also Table 2).
Therefore, we conclude that we detect molecular hydrogen absorption
from the $J=0$ to $J=4$ states, associated with the damped Ly$\alpha$ system at $z=1.7765$.

In Fig. 2, we mark the expected positions of metal
lines from all the known absorbers, including the Milky Way interstellar medium,
on the observed spectrum (solid line) with $1\:\sigma$ errors (dotted line).
The atomic line list was taken from Table 2 of Prochaska {\it et al.} (2001).
Also marked on Fig. 2 are all the ${\rm H}_2$ lines from $J=0$ to $4$ in the Lyman
and Werner bands for the $z=1.7765$ absorber.

\begin{figure*}
\figurenum{2a}
\includegraphics[angle=90,scale=0.6]{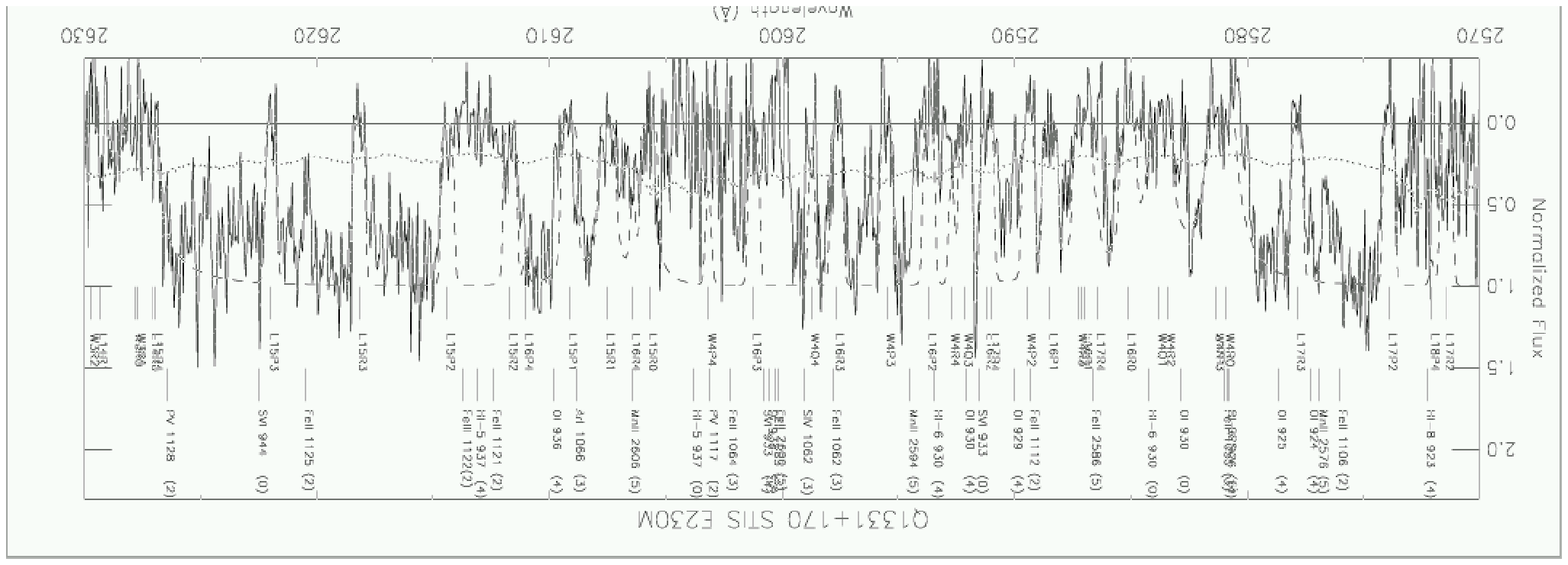}
\caption{The STIS E230M spectrum of Q1331+170 (solid line) with line identifications
and $1\:\sigma$ errors (dotted line).
The molecular hydrogen lines from $J=0$ to $4$ in the Lyman
and Werner bands, as well as the expected positions of metal lines from known
metal absorbers are marked. Numbers in the parenthesis following the
metal line identifications indicate different absorption systems: 0 - the damped
Ly$\alpha$ absorber at $z=1.7765$, 1 - the absorber at $z=0.7441$, 2 - the absorber
at $z=1.3284$, 3 - the absorber at $z=1.4458$, 4 - the absorber at $z=1.7864$, and
5 - the Milky Way. The dashed line gives the ${\rm H}_2$ synthetic spectrum,
calculated with the best-fit parameters in Table 3.}
\end{figure*}

\begin{figure*}
\figurenum{2b}
\includegraphics[angle=90,scale=0.6]{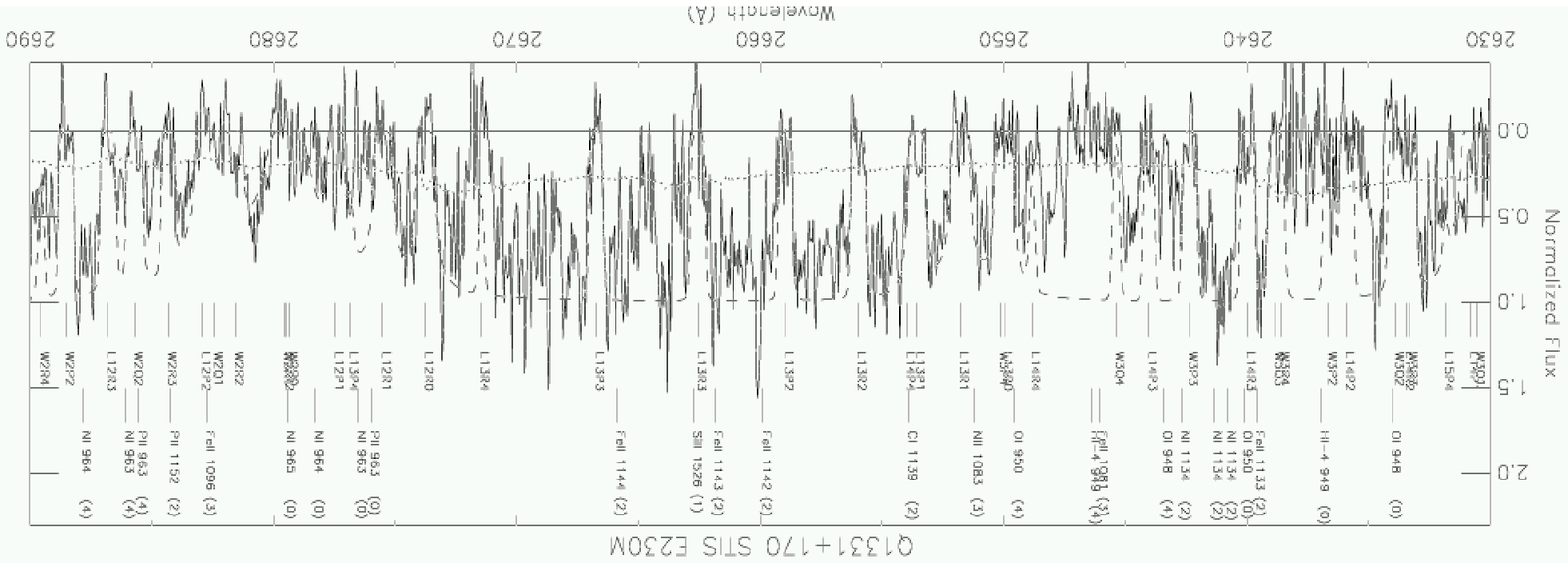}
\caption{Same as Fig. 2a.}
\end{figure*}

\begin{figure*}
\figurenum{2c}
\includegraphics[angle=90,scale=0.6]{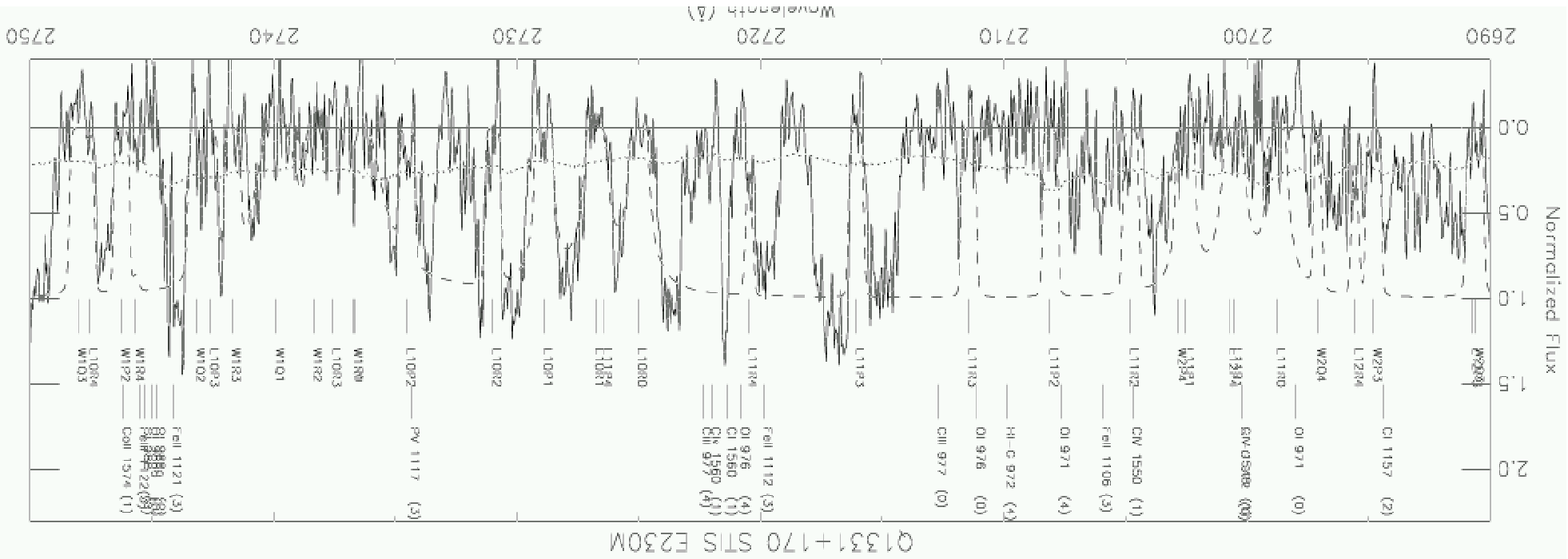}
\caption{Same as Fig. 2a.}
\end{figure*}

\begin{figure*}
\figurenum{2d}
\includegraphics[angle=90,scale=0.6]{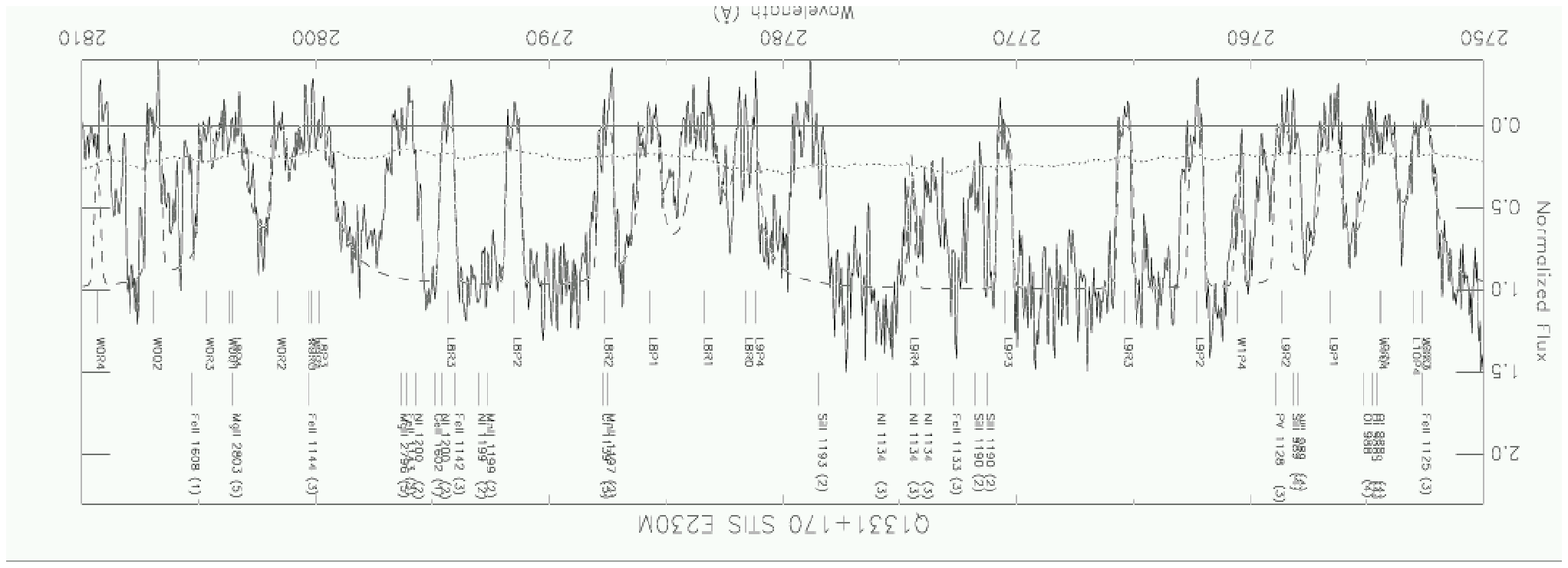}
\caption{Same as Fig. 2a.}
\end{figure*}

\begin{figure*}
\figurenum{2e}
\includegraphics[angle=90,scale=0.6]{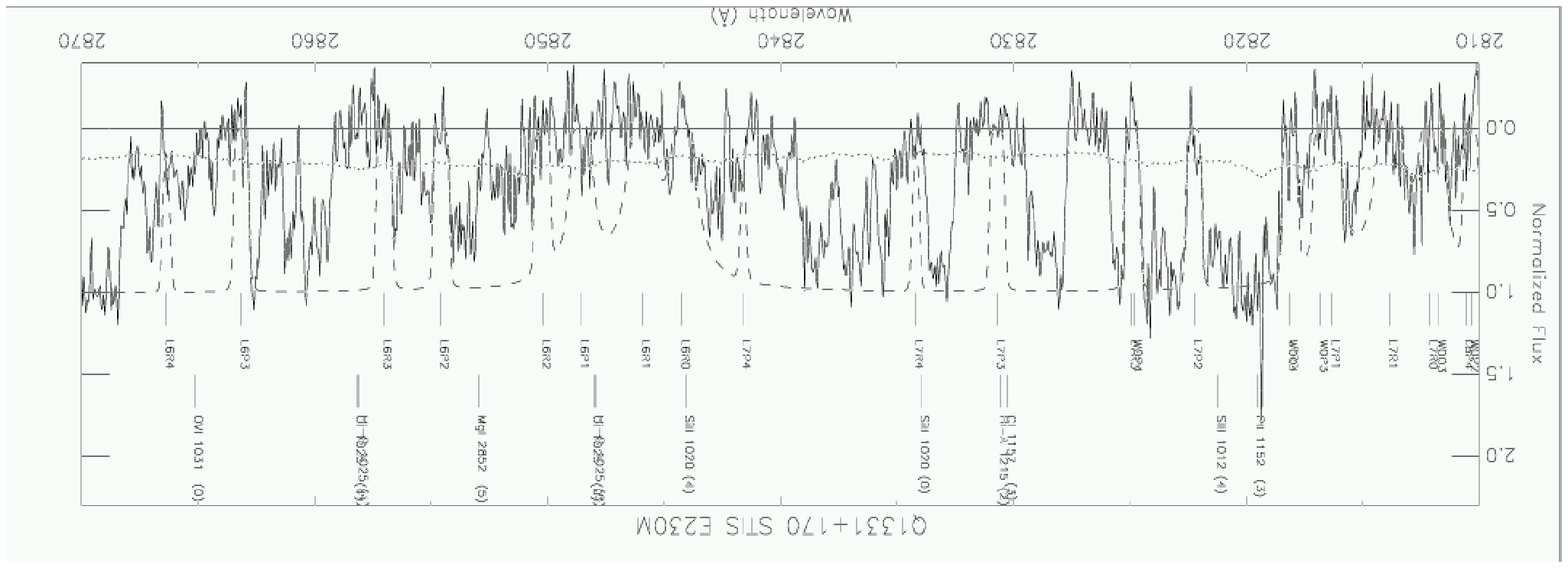}
\caption{Same as Fig. 2a.}
\end{figure*}

\begin{figure*}
\figurenum{2f}
\includegraphics[angle=90,scale=0.6]{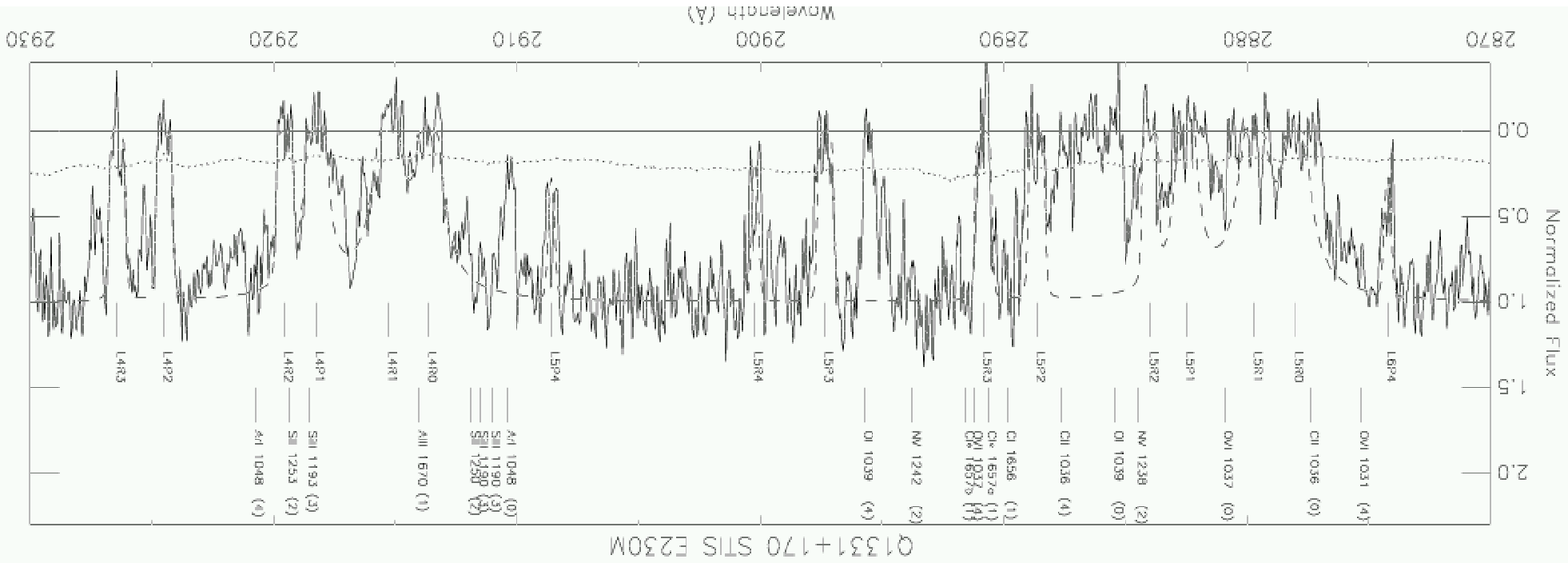}
\caption{Same as Fig. 2a.}
\end{figure*}

\begin{figure*}
\figurenum{2g}
\includegraphics[angle=90,scale=0.6]{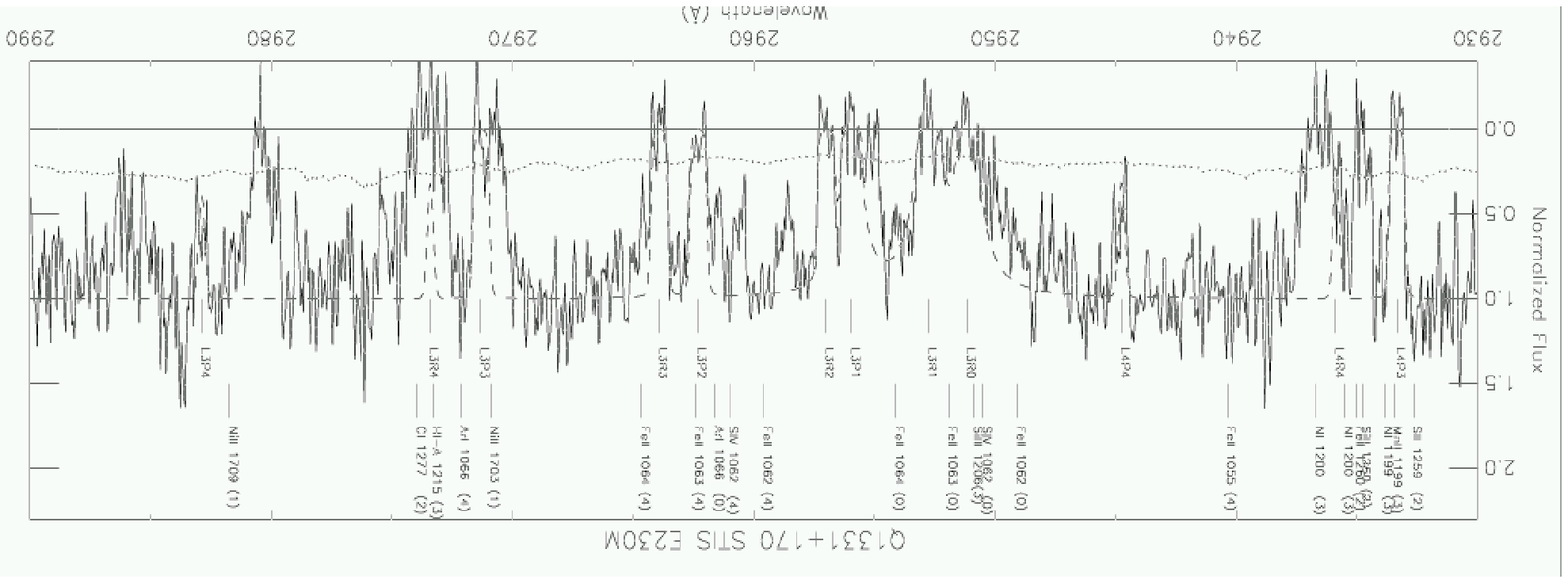}
\caption{Same as Fig. 2a.}
\end{figure*}

\begin{figure*}
\figurenum{2h}
\includegraphics[angle=90,scale=0.6]{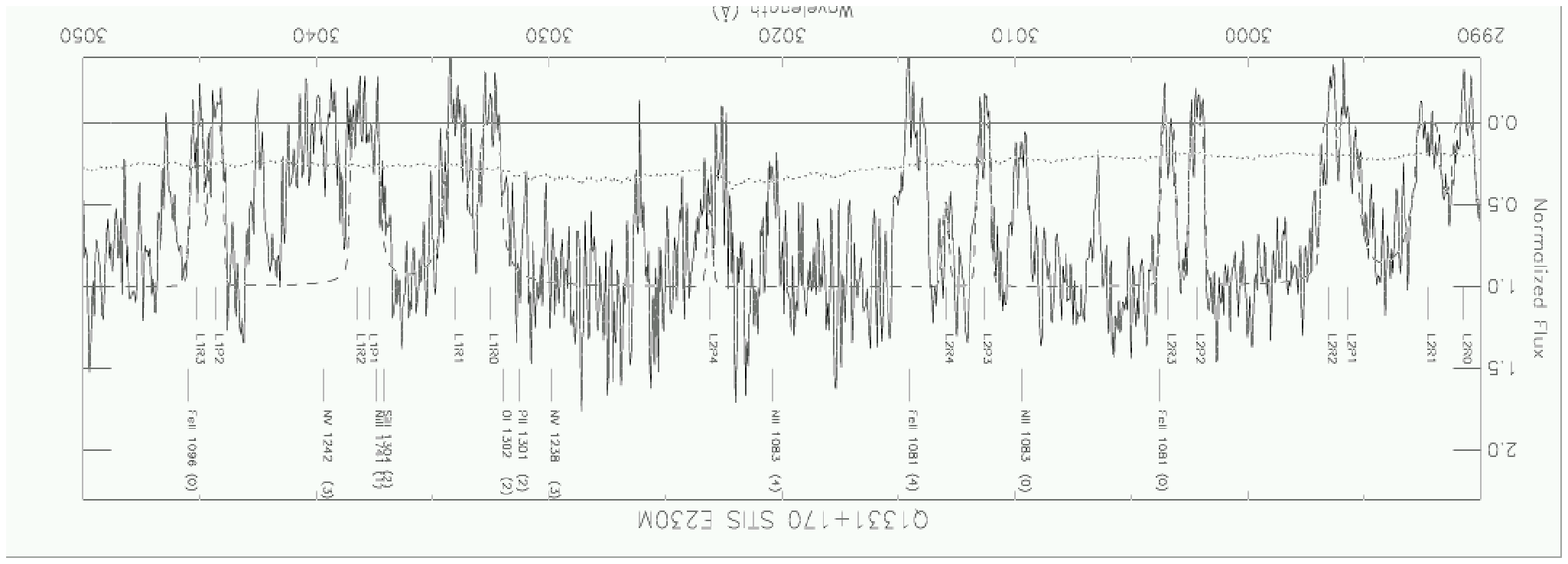}
\caption{Same as Fig. 2a.}
\end{figure*}

\subsection{Voigt profile fitting of molecular hydrogen lines}

We used VPFIT (version 5) to simultaneously fit Voigt profiles for all the ${\rm H}_2$ lines. 
All lines with $J=0,\:1,\:2,\:3$ and $4$ were fit simultaneously,
assuming one $b$-parameter and redshift for all lines and different column densities
for each $J$ state.
These free parameters were adjusted independently to minimize the $\chi^2$
value of the overall fitting. We adopted the STIS instrument profile for the echelle E230M
and $0.2^{\arcsec}\times 0.2^{\arcsec}$ aperture.
Those $H_2$ lines with large reduced $\chi^2$ deviations ($\chi_\nu^2>1.2$)
were excluded, to avoid possible cases in which ${\rm H}_2$ absorption may be
contaminated by Ly$\alpha$ forest lines.
This procedure was iterated until no more lines were rejected.
Our final model includes 20 different regions with 26 molecular hydrogen lines for simultaneous
fitting (two $J=0$ lines, three $J=1$ lines, seven $J=2$ lines, twelve $J=3$ lines and two $J=4$ lines).
We also note that the physical parameters determined from Voigt profile fitting
(see below) are insensitive to our adopted maximum reduced $\chi^2$ deviation of $1.2$.

\begin{figure*}
\figurenum{3}
\epsscale{1.0}
\plotone{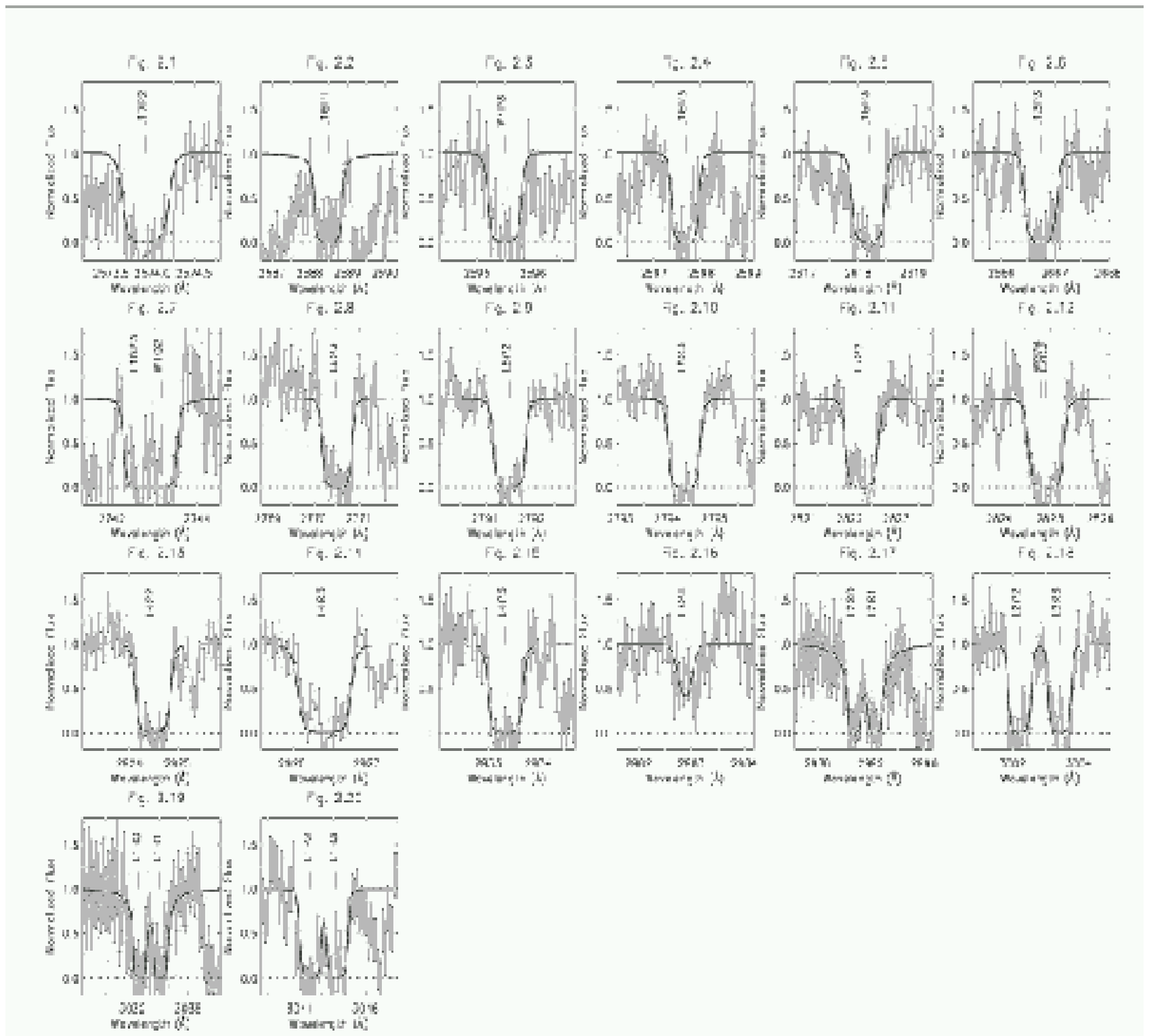}
\caption{The best-fit Voigt profiles for the 26 ${\rm H}_2$ lines used in our
simultaneous fitting (solid lines), overplotted on the observed
profiles with $1\:\sigma$ errors.}
\end{figure*}

Table 2 provides the results for all the ${\rm H}_2$ lines included in the
final fitting. Rest-frame wavelengths, $f$-values and damping parameters $\Gamma$
are taken from Abgrall {\it et al.}
(1993a, 1993b). The sixth column gives the reduced $\chi^2$ value for all
the 20 fitted regions.
We show the best-fit Voigt profiles for different ${\rm H}_2$ lines in Fig. 3 (solid lines),
superposed on the observed spectrum with $1\:\sigma$ errors. The Voigt
profiles for all the ${\rm H}_2$ lines with $J=0$ to $4$ in the Lyman and Werner bands
are also shown by the dashed line in Fig. 2.

\begin{deluxetable}{cccccc}
\tablecaption{${\rm H}_2$ absorption lines associated with the DLA toward Q1331+170\label{tb2}}
\tablewidth{0pt}
\tablehead{
\colhead{No.}&\colhead{Identification}&\colhead{$\lambda_{\rm rest}$ (\AA)}&
\colhead{$f$ ($10^{-2}$)}&\colhead{$\Gamma$ ($10^9\:{\rm s^{-1}}$)}&
\colhead{${\chi_{\nu}^2}$}}
\startdata
 1&L17P2& 927.02&0.233&0.804&1.01\\
 2&L16P1& 932.27&0.240&0.835&0.32\\
 3& W4P3& 934.79&0.709&1.101&0.90\\
 4&L16R3& 935.57&0.657&0.830&0.80\\
 5&L15R3& 942.96&0.574&0.862&1.06\\
 6&L13P3& 960.45&0.493&0.939&0.73\\
 7&L10P3& 987.77&0.893&1.070&1.10\\
  & W1Q2& 987.97&3.650&1.157&    \\
 8& L9P3& 997.82&0.876&1.120&0.76\\
 9& L8P2&1005.39&0.991&1.177&0.46\\
10& L8R3&1006.41&1.572&1.167&0.73\\
11& L7P2&1016.46&1.023&1.236&0.72\\
12& W0P4&1017.38&0.995&1.178&1.10\\
  & L7R3&1017.42&1.838&1.224&    \\
13& L4P2&1053.28&0.902&1.448&0.70\\
14& L4R3&1053.98&1.336&1.434&0.85\\
15& L4P3&1056.47&0.956&1.445&0.59\\
16& L3P4&1074.31&0.774&1.525&0.98\\
17& L2R0&1077.14&1.166&1.630&0.62\\
  & L2R1&1077.70&0.769&1.626&    \\
18& L2P2&1081.27&0.470&1.630&0.93\\
  & L2R3&1081.71&0.636&1.613&    \\
19& L1R0&1092.19&0.578&1.738&0.88\\
  & L1R1&1092.73&0.378&1.734&    \\
20& L1P2&1096.44&0.237&1.738&0.94\\
  & L1R3&1096.73&0.305&1.719&    \\
\enddata
\end{deluxetable}

The best-fit results of ${\rm H}_2$ column densities, $b$-parameter and redshift
are listed in Table 3. The $1\:\sigma$ errors were derived from $\chi^2$ fitting by VPFIT,
and are mainly dependent on photon statistics.
The total molecular hydrogen column density
is $N_{\rm H_2}=(4.45\pm 0.36)\times 10^{19}\:{\rm cm^{-2}}$. We can
place an upper limit on the $J=5$ column density
as $N_{J=5}\le 6.3\times 10^{13}\:{\rm cm^{-2}}$, at $2\:\sigma$ significance level.
The $1\:\sigma$ error in redshift given in Table 3 corresponds to the uncertainty in
line centroiding, which is equivalent with $0.9\:{\rm km\:s^{-1}}$ in velocity space. 
We mentioned in Sec. 2 that the velocity uncertainty caused by wavelength calibration
is $0.8\:{\rm km\:s^{-1}}$. Therefore the total $1\:\sigma$ error is $1.2\:{\rm km\:s^{-1}}$,
which is about 10\% of the spectral resolution.

\begin{deluxetable}{ccc}
\tablecaption{${\rm H}_2$ column densities$^a$, $b$-parameter and redshift\label{tb3}}
\tablewidth{0pt}
\tablehead{
\colhead{Parameter}&\colhead{Measured value}&\colhead{$1\:\sigma$ error}}
\startdata
$z$&1.776553&0.000003\\
$b$ (${\rm km\:s^{-1}}$)&13.85&0.52\\
$\log(N_{J=0})$&19.144&0.066\\
$\log(N_{J=1})$&19.467&0.043\\
$\log(N_{J=2})$&17.901&0.187\\
$\log(N_{J=3})$&17.674&0.184\\
$\log(N_{J=4})$&15.101&0.125\\
$\log(N_{\rm total})$&19.648&0.035\\
\enddata
\tablenotetext{a}{Column densities are in units of cm$^{-2}$.}
\end{deluxetable}

The molecular hydrogen fraction is defined as
\begin{equation}
f_{\rm H_2}=\frac{2N_{\rm H_2}}{N_{\rm HI}+2N_{\rm H_2}}.
\end{equation}
\noindent
Adopting a neutral hydrogen column density of $N_{\rm HI}=(1.50\pm 0.14)\times 10^{21}\:{\rm cm^{-2}}$
(Prochaska \& Wolfe 1999),
we derived a molecular hydrogen fraction of $f_{\rm H_2}=(5.6\pm 0.7)\%$.
This is the largest value reported so far in any redshifted damped Ly$\alpha$ system.
The profiles of the Lyman series lines are shown in Fig. 1, assuming
$N_{\rm HI}=1.50\times 10^{21}\:{\rm cm^{-2}}$ and $b=19.6\:{\rm km\:s^{-1}}$.
The $b$-parameter is directly scaled from the $b$-parameter determined from
Voigt profile fitting of molecular hydrogen lines, taking into account the difference
in particle mass between the two species.
The synthetic line profiles are in broad agreement with the observations.
The fact that H~I lines are heavily blended with other ${\rm H}_2$ lines makes Voigt profile fitting
to derive $N_{\rm HI}$ difficult. Therefore we simply adopt the neutral hydrogen column
density from previous work (Prochaska \& Wolfe 1999).

\subsection{Uncertainties due to curve-of-growth and continuum fitting}

The uncertainties in the measured ${\rm H}_2$ column densities at different $J$ states
depend on the location of the corresponding lines on the curve of growth.
We show in Fig. 4 the curve of growth for the 26 ${\rm H}_2$ lines used in our simultaneous
Voigt profile fitting. Note that the equivalent widths are calculated with the
best-fit parameters shown in Table 3, since some of the lines are blended and
therefore not appropriate for direct measurements of equivalent widths from the observed
spectrum. Lines with different $J$ values are marked with different symbols.
The background dots in Fig. 4 bracket the full range of the curve of growth
for ${\rm H}_2$ lines, taking into account the variations in the atomic data
(Abgrall {\it et al.} 1993a, 1993b). The figure shows that
the $J=0$, $1$ lines lie on the damped part of the curve of growth, while $J=2$, $3$ lines
on the flat part, and $J=4$ lines on the linear part of the curve of growth.
Since the determination of
column densities is very sensitive to the choice of $b$-parameter on the flat
part of the curve of growth, the uncertainties in $N_{J=2}$ and $N_{J=3}$ are
relatively larger than the column densities at lower $J$ states.
Although $J=4$ lines are on the linear part of the curve of growth,
the $1\:\sigma$ error in $N_{J=4}$ is also relatively large due to the
low signal-to-noise ratio for these weaker lines. 

\begin{figure}
\figurenum{4}
\epsscale{1.0}
\plotone{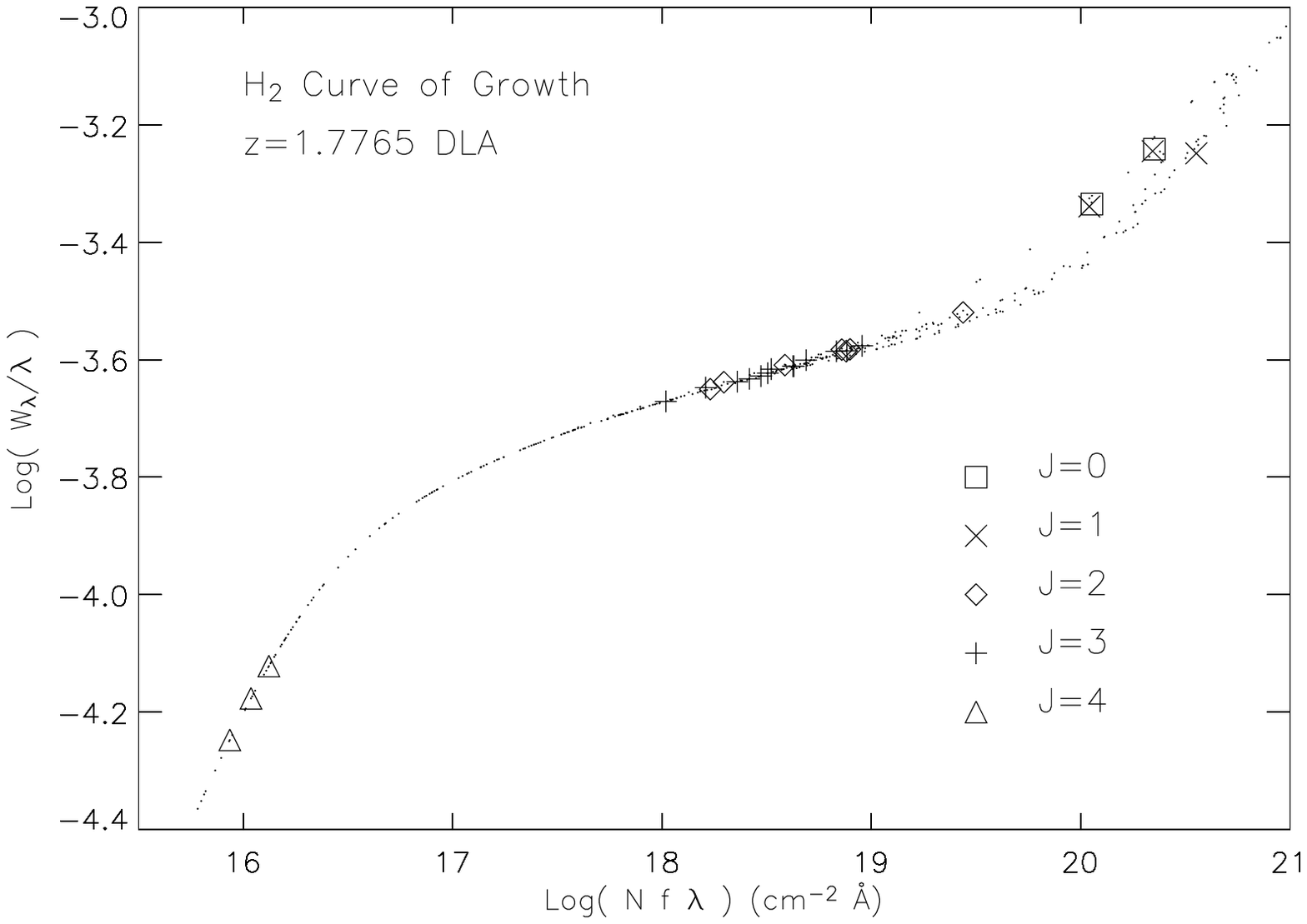}
\caption{The curve of growth for ${\rm H}_2$ lines identified in the spectrum
of Q1331+170. Different symbols stand for transitions from different $J$ states.
The background dots bracket the full range of the curve of growth for
${\rm H}_2$ lines, taking into account variations in the atomic
data. The $J=0$, $1$ lines are on the damped part of the curve of growth, the
$J=2$, $3$ lines on the flat part, and the $J=4$ lines on the linear part.}
\end{figure}

Uncertainties in column density are also likely introduced by uncertainties in continuum fitting.
To investigate this effect, we scaled our continuum by a factor ranging from 25\%
lower to 25\% higher than the best-fit continuum.
Then we did simultaneous Voigt profile fitting
in the same way as described in Sec. 3.2. The results are shown in Fig. 5, for the measured
column densities at different $J$ states. We find that the errors in column density at all
$J$ states caused by $1\:\sigma$ uncertainties in continuum fit
are considerably smaller than the $1\:\sigma$ error in column density derived from Voigt profile fitting.
In the extreme case that if our best-fit continuum
is under-estimated by a factor of 25\% (which can be excluded by 98\% confidence level,
see Sec. 3.2), the corresponding uncertainties in ${\rm H}_2$ column
densities at $J=2$ and $3$ can be relatively large. Fortunately, the physical state in the ${\rm H}_2$
absorber is not sensitive to ${\rm N}_{J=2,\:3}$ (see Sec. 4.1 for more details). Therefore
we conclude that our final results are not critically dependent on the uncertainties in continuum fitting.

\begin{figure}
\figurenum{5}
\epsscale{1.0}
\plotone{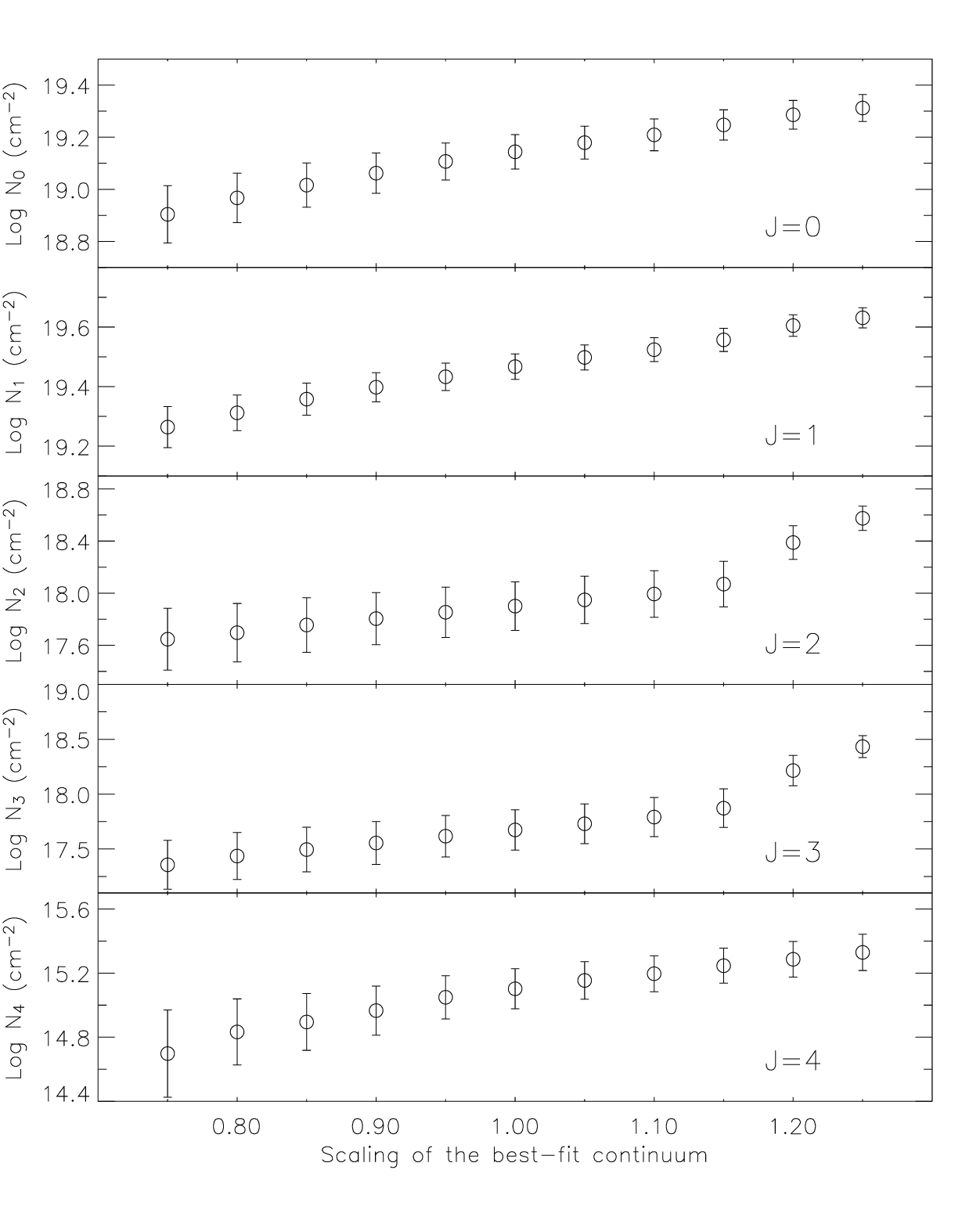}
\caption{The effect of continuum fitting on the measured column densities
of ${\rm H}_2$. The abscissa shows the scaling factor applied to the best-fit
continuum shown in Fig. 1. The best-fit continuum corresponds to a scaling
factor of 1.0.}
\end{figure}

\subsection{Excitation temperature}

Once the ${\rm H}_2$ column densities at different $J$ states are determined, we can calculate
the excitation temperature for each $J\ge 1$ state by the Boltzmann law
\begin{equation}
\frac{N_J}{N_0}=\frac{g_J}{g_0}\exp{[-\frac{\epsilon_J-\epsilon_0}{kT_{\rm ex}}]},
\end{equation}
\noindent
where $\epsilon_J-\epsilon_0$ is the excitation energy of level $J$ relative to the ground state,
$k$ is the Boltzmann constant,
the degeneracy is $g_J=(2I+1)\times (2J+1)$, and $I$ is the nuclear spin (0 for even $J$
and 1 for odd $J$). Fig. 6 shows the plot of $\log{\frac{N_{J}}{g_J}}$ for each $J$ state
as a function of its excitation potential, with $1\:\sigma$ errors.
The populations of the five detected $J$ states can be well
fitted by a single temperature Boltzmann distribution, with $\chi_\nu^2=0.20$.
The one-temperature model is shown by the solid line in Fig. 6, with
a best-fit excitation temperature of $T_{\rm ex}=(152\pm 10)\:{\rm K}$.
The upper limit put on $N_{J=5}$ is also consistent with this model.
Increasing or decreasing the best-fit continuum by 15\% results in a similar
excitation model with the best-fit excitation temperature of
$T_{\rm ex}=(153\pm 10)\:{\rm K}$ or $T_{\rm ex}=(150\pm 10)\:{\rm K}$,
respectively. The insensitivity of $T_{\rm ex}$ to the continuum fitting partly results from the
logarithmic dependence of $T_{\rm ex}$ on the ratio of column densities.

\begin{figure}
\figurenum{6}
\epsscale{1.0}
\plotone{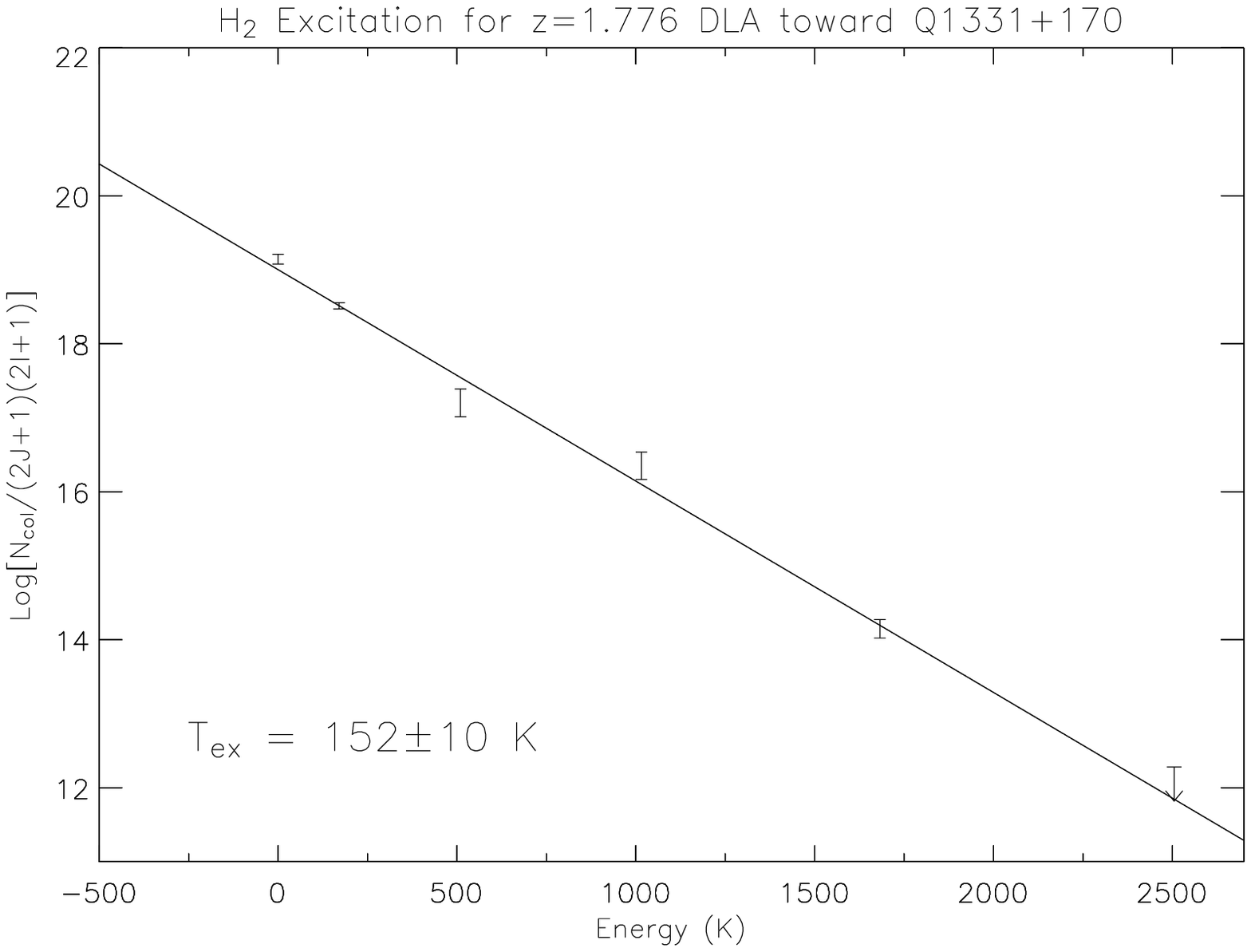}
\caption{Distribution of molecular hydrogen at different $J$ levels.
The abscissa gives the relative
energy for each $J$ state relative to $J=0$, expressed in units of Kelvin;
the ordinate gives the column density divided by the degeneracy of
the corresponding state, with $1\:\sigma$ errors.
The population of ${\rm H}_2$ at different $J$ states
can be fitted by a one-temperature Boltzmann distribution,
with $T_{\rm ex}\approx 152\:{\rm K}$.}
\end{figure}

A unique temperature for rotational excitation of ${\rm H}_2$ is not characteristic
of Galactic ${\rm H}_2$ clouds, and the populations of rotational states of ${\rm H}_2$ often
require at least two different excitation temperatures (e.g. Spitzer \& Cochran 1973).
$T_{\rm ex}$ for the low $J$ states approximates the kinetic temperature of the gas,
while $T_{\rm ex}$ for high $J$ states results from UV pumping
(Spitzer {\it et al.} 1974). For the damped Ly$\alpha$ absorber of Q1331+170,
a unique excitation temperature for all states from $J=0$ to $4$
indicates that the rotational populations of ${\rm H}_2$ are probably dominated by collisional
excitation in a gas with kinetic temperature around $150\:{\rm K}$.
Moreover, the local UV radiation field does not significantly populate the high $J$ states
of ${\rm H}_2$ molecules. We quantify the UV radiation field below (see Sec. 4.1).

\section{Discussion}

\subsection{UV radiation field}

A simple analysis on the rate equation of the $J=4$ state can yield an estimate of the
UV photo-absorption rate of ${\rm H}_2$.
Collisions do not provide an important population source
for $J\ge 4$ states (e.g. Browning {\it et al.} 2003), unless the gas phase temperature
is comparable with or higher than
$T\approx\frac{\epsilon_{J\ge 4}-\epsilon_0}{k}\gtrsim 1700\:{\rm K}$.
However, such a high temperature can be excluded for the damped Ly$\alpha$
absorber toward Q1331+170, either by the distribution of
${\rm H}_2$ at different $J$ states (see Fig. 6), or by the observation of 21 cm absorption
line (Wolfe \& Davis 1979, Chengalur \& Kanekar 2000).

Previous work shows that the $J=4$ state is mainly populated
by direct formation into this level and pumping from $J=0$, while the dominant
depopulation mechanism is spontaneous decay (e.g., Dalgarno \& Wright 1972, Jura 1975).
Assuming a steady state, these effects can be combined into the following equation
\begin{equation}
p_{4,0}R_{\rm abs}n({\rm H}_2,J=0)+0.19R_{\rm dust}n({\rm HI})n({\rm H})=A_{42}n({\rm H}_2,J=4),
\end{equation}
\noindent
where $R_{\rm abs}$ is the photo-absorption rate,
$n({\rm HI})$ is the neutral hydrogen number density, $n({\rm H})\approx n({\rm HI})+2n({\rm H}_2)$
is the total hydrogen number density, $R_{\rm dust}$ is the ${\rm H}_2$ formation
rate on the surface of dust grains, $p_{4,0}=0.26$ is the UV pumping efficiency from $J=0$ to $J=4$ (Jura 1975),
and $A_{42}=2.75\times 10^{-9}\:{\rm s^{-1}}$ is the spontaneous transition probability
from $J=4$ to $J=2$ (Wolniewicz {\it et al.} 1998).
For simplicity, we assume 19\% of all ${\rm H}_2$ formation results in the population
at $J=4$ (Jura 1975), although more recent work suggests that this fraction may
vary for different types of dust grains (Takahashi \& Uehara 2001).

The equilibrium between the formation and photo-dissociation of ${\rm H}_2$
can be written as
\begin{equation}
R_{\rm dust}n({\rm HI})n({\rm H})\approx 0.11R_{\rm abs}n({\rm H}_2),
\end{equation}
\noindent
where we assume that approximately 11\% of photo-absorption
leads to photo-dissociation (Jura 1974).
Combining Eqn. 4 and 5, we get $R_{\rm abs}=(7.6\pm 2.4)\times 10^{-13}\:{\rm s^{-1}}$.
We note that the photo-absorption rate derived this way is independent of the
neutral hydrogen density, since the relevant term $Rn({\rm HI})n({\rm H})$ cancels
from Eqn. 4 and 5. Therefore $R_{\rm abs}$ is independent of the adopted H~I column density
(see Sec. 3.5).

The photo-absorption rate, $R_{\rm abs}$ derived above is based on the ${\rm H}_2$
population at the $J=4$ state. Since $J=4$ lines are optically thin
(see Sec. 3.3), the photo-absorption
rate within the ${\rm H}_2$ cloud should be approximately the same as that
outside the cloud. Therefore $R_{\rm abs}$ is also characteristic
of the UV background intensity in the absorber's environment.
With this in mind, we use $R_{\rm abs}$ derived above to estimate the ambient UV radiation field
at $1000\:{\rm \AA}$ from the expression,
\begin{equation}
R_{\rm abs}=4.0\times 10^{-10}S_{\rm shield}\frac{J_{1000\:{\rm \AA}}}{J_{1000\:{\rm \AA},\odot}}\:{\rm s^{-1}},
\end{equation}
\noindent where $J_{1000\:{\rm \AA},\odot}\approx 3.2\times 10^{-20}\:{\rm ergs\:s^{-1}\:cm^{-2}\:Hz^{-1}\:Sr^{-1}}$
is the UV intensity at $1000\:{\rm \AA}$ in the solar vicinity (Hirashita \& Ferrara 2005), 
and $S_{\rm shield}$ corrects for the the ${\rm H}_2$ self-shielding and/or dust extinction.
Since the $J=4$ lines are optically thin, the self-shielding of ${\rm H}_2$ can be reasonably ignored,
and we estimate the importance of dust extinction following Hirashita \& Ferrara (2005), i.e.
\begin{equation}
S_{\rm shield}\approx \exp{[-0.879\kappa(\frac{N_{\rm HI}}{10^{21}\:{\rm cm^{-2}}})]},
\end{equation}
\noindent where $\kappa$ is the dust-to-gas ratio, defined as the ratio of the extinction
optical depth to neutral hydrogen column density in units of $10^{21}\:{\rm cm^{-2}}$ (Pei {\it et al.} 1991).
Vladilo (1998) showed that $\kappa$ can be estimated by
\begin{equation}
\kappa=1.7\times 10^{[{\rm Zn/H}]}(1-10^{[{\rm Fe/Zn}]}).
\end{equation}
Assuming that ZnII and FeII are the dominant ionization states for these two species,
we adopt $[{\rm Fe/H}]=-1.22$ and $[{\rm Fe/Zn}]=-0.87$ for the Q1331+170 DLA from Prochaska \& Wolfe (1999).
This gives $\kappa\approx 0.088$, and therefore $S_{\rm shield}\approx 0.89$, with our adopted H~I column density
(see Sec. 3.2). Inserting the above values into Eqn. 6, we get
\begin{equation}
\frac{J_{1000\:{\rm \AA}}}{J_{1000\:{\rm \AA},\odot}}\approx (2.1\pm 0.7)\times 10^{-3},
\end{equation}
\noindent i.e. the ambient UV radiation field of the ${\rm H}_2$ cloud associated with DLA 1331+170
is about three orders of magnitude weaker than that in the Solar vicinity.
Assuming the spectrum of the UV radiation field follows $J_{\nu}\propto \nu^{-0.5}$,
we estimate the UV intensity at the Lyman limit as
\begin{equation}
J_{912\:{\rm \AA}}\approx (7.1\pm 2.3)\times 10^{-23}\:{\rm ergs\:s^{-1}\:cm^{-2}\:Hz^{-1}\:Sr^{-1}}.
\end{equation}
\noindent This value is consistent with the mean metagalactic UV background intensity of
$J_{912\:{\rm \AA}}\approx 7.6^{+9.4}_{-3.0}\times 10^{-23}\:{\rm ergs\:s^{-1}\:cm^{-2}\:Hz^{-1}\:Sr^{-1}}$
at similar redshifts determined from proximity effect in the Ly$\alpha$ forest (Scott {\it et al.} 2002),
implying that the ambient UV radiation of the ${\rm H}_2$ absorber may originate externally from the cloud.
If instead the UV field is stellar in origin, $J_{912\:{\rm \AA}}$ would be lower than the metagalactic
value, so our assumption of $J_{\nu}\propto \nu^{-0.5}$ gives a reasonable upper limit to the ionizing
radiation field.
Based on the above calculations, the existence of any significant internal source of UV emission is not allowed, 
suggesting negligible star formation activity associated with this ${\rm H}_2$ absorber.
Finally, we emphasize that our estimate of $J_{912\:{\rm \AA}}$ is based on the assumption
of QSO-only spectral shape for the UV background intensity. This is a reasonable assumption if
star-forming galaxies do not make significant contribution to the metagalactic UV radiation field
(e.g. Leithere {\it et al.} 1995, Hurwitz {\it et al.} 1997, Deharveng {\it et al.} 2001,
Giallongo {\it et al.} 2002, Fern\'andez-Soto {\it et al.} 2003).

The ambient UV intensity for DLA 1331+170 is much smaller than that in other DLAs,
which is usually close to the typical Galactic value
(e.g. Ge \& Bechtold 1997, Levshakov {\it et al.} 2002, Ledoux {\it et al.} 2002).
Hirashita \& Ferrara (2005) estimated that the probable range of radiation field is
$0.5\lesssim \log{(\frac{J_{1000\:{\rm \AA}}}{J_{1000\:{\rm \AA},\odot}})}\lesssim 1.5$,
for a sample of DLAs which have been previously searched for ${\rm H}_2$ absorption
(Ledoux {\it et al.} 2003). The associated surface star formation rate (SFR) is
$5\times 10^{-3}\:{\rm M_\odot\:yr^{-1}\:kpc^{-2}}\lesssim\Sigma_{\rm SFR}\lesssim 5\times 10^{-2}\:{\rm M_\odot\:yr^{-1}\:kpc^{-2}}$
(Hirashita \& Ferrara 2005, see also Wolfe {\it et al.} 2003a, 2003b).
Also, we note that Reimers {\it et al.} (2003) reported the detection of ${\rm H}_2$
associated with a sub-DLA system, with local UV radiation field
inferred to be $\sim 300$ times stronger than the mean Galactic value.
Although most DLAs may be associated with considerable
on-going star formation, the DLA toward Q1331+170 at least represents a special system,
which is in a stage either before the onset of considerable star formation, or after
star forming activities have ceased.

For comparison, we note that the C~II* 1336${\rm \AA}$ absorption feature is useful for probing the ambient UV intensity
near $1500\:{\rm \AA}$ (Wolfe {\it et al.} 2003a, 2003b). Thus it is interesting to compare
the UV radiation field inferred by C~II* with that determined from ${\rm H}_2$.
Using the C~II* absorption line as a diagnostic of the UV radiation field follows the idea that
$[{\rm CII}]\:158\mu m$ emission is the dominant coolant in DLAs, and is balanced by the grain photoelectric
heating in a steady state. Since the grain photoelectric heating rate is proportional to the ambient UV intensity,
a measurement of the strength of C~II* 1336${\rm \AA}$ absorption provides an estimate of the UV radiation field
at $1500\:{\rm \AA}$. Following Wolfe {\it et al.} (2003a), we express the heating rate per H atom as
\begin{equation}
\Gamma=10^{-24}\kappa\epsilon (\frac{J_{1500\:{\rm \AA}}}{10^{-19}\:{\rm ergs\:s^{-1}\:cm^{-2}\:Hz^{-1}\:Sr^{-1}}})\:{\rm ergs\:s^{-1}\:H^{-1}},
\end{equation}
\noindent where $\kappa=0.088$ is the dust-to-gas ratio determined from Eqn. 8, $\epsilon\approx 0.049$ is the
heating efficiency (Bakes \& Tielens 1994, Wolfire {\it et al.} 1995), and $J_{1500 {\rm \AA}}$ is the UV intensity at $1500\:{\rm \AA}$.
For the DLA towards Q1331+170, the C~II* 1336${\rm \AA}$ absorption line strength implies
a cooling rate of $l_{\rm c}\approx 2.24\times 10^{-27}\:{\rm ergs\:s^{-1}\:H^{-1}}$
(Wolfe {\it et al.} 2003a). Equating $\Gamma$ and $l_{\rm c}$ implies
$J_{\rm 1500\:{\rm \AA}}\approx 5.4\times 10^{-20}\:{\rm ergs\:cm^{-2}\:s^{-1}\:Hz^{-1}\:Sr^{-1}}$,
not too far from the mean Galactic value and much higher than the UV radiation field inferred from ${\rm H}_2$ absorption.
This discrepancy is not expected, since the large difference in the UV intensity determined from
${\rm H}_2$ and C~II* is difficult to be interpreted by any local sources of absorption.
Note that ${\rm H}_2$ and C~II* absorption probe UV intensity near $\sim 1000\:{\rm \AA}$ and
$\sim 1500\:{\rm \AA}$, respectively, which are both above the Lyman limit. This indicates that photoionization
of H~I is not relevant. We also mentioned above that $J_{\rm 1000\:{\rm \AA}}$ determined from ${\rm H}_2$
absorption is characteristic of the UV radiation field outside the ${\rm H}_2$ cloud, therefore
molecular hydrogen absorption in the Lyman and Werner bands does not account for the discrepancy either.
However, the very low ambient UV intensity determined for the ${\rm H}_2$ absorber is reasonable,
otherwise the residual excitation temperature inferred from C~I/C~I* absorption would be far
below the cosmic microwave background (CMB) temperature predicted by the standard cosmology (see Sec. 4.7).

\subsection{The structure of the ${\rm H}_2$ absorber}

To study the structure of the ${\rm H}_2$ cloud associated with the DLA of Q1331+170, 
we use CLOUDY (version c9400)
to construct a grid of models with different total hydrogen number density
$n({\rm H})\approx n({\rm HI})+2n({\rm H}_2)$, and examine
which model provides the best match with the observed ${\rm H}_2$ column density.
We adopt a metallicity of $[{\rm Zn}/{\rm H}]=-1.22$
(Prochaska \& Wolfe 1999) for all the elements in the absorber.
To take into account differential dust depletion, we determine the dust-to-gas ratio
$\kappa$ according to Eqn. 8.
The ambient UV intensity is characterized by the metagalactic UV radiation field,
assuming a QSO-like spectral shape, as implied by results in Sec. 4.1.
For simplicity, we assume plane parallel geometry with a uniform total hydrogen
number density. The cloud is illuminated from both sides in our model,
and we run the calculation until the neutral hydrogen column density reaches
the observed value of $N({\rm HI})=1.5\times 10^{21}\:{\rm cm^{-2}}$. 

\begin{figure}
\figurenum{7}
\epsscale{1.0}
\plotone{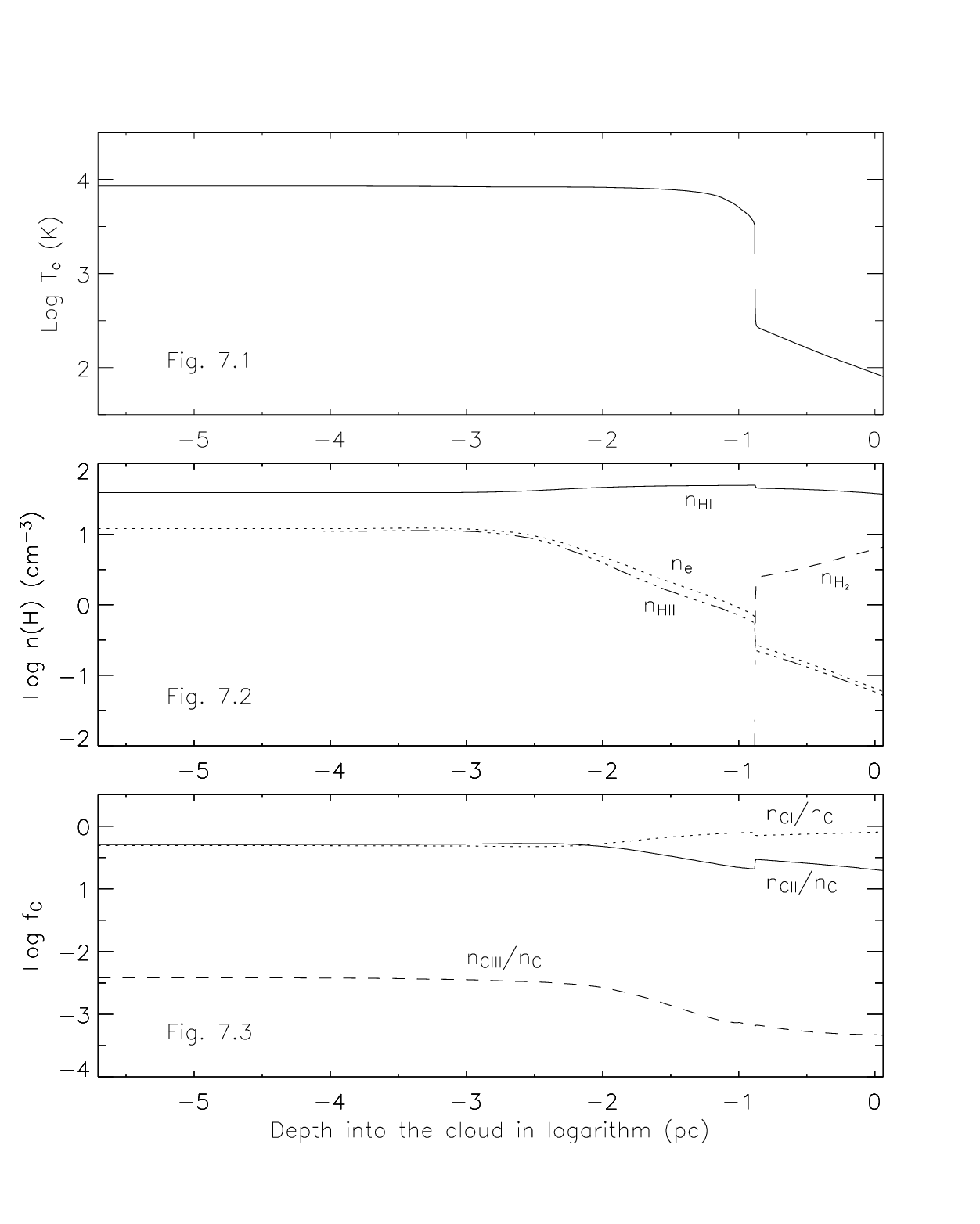}
\caption{Model calculations of the thermal and ionization structure of the ${\rm H}_2$
cloud (see Sec. 4.2). The electron temperature (upper
panel), the number densities of H~I, H~II, ${\rm H}_2$, electrons (middle panel),
and the fraction of different species of carbon (lower panel)
are each plotted as a function of the depth into the cloud.
Note that only half of the cloud is shown in the figure ($\sim 1.2\:{\rm kpc}$).}
\end{figure}

The model with a total hydrogen number density, $n({\rm H})\approx n({\rm HI})+2n({\rm H}_2)\approx 0.22$,
gives the best match to the observed ${\rm H}_2$ fraction for the $z=1.7765$ DLA system.
We show in Fig. 7.1 the thermal structure of the cloud, with the electron temperature determined
from the balance between cooling and heating. In Fig. 7.2, we give the structure of the cloud
traced by H~I, H~II, ${\rm H}_2$ and $e$. We find that the neutral hydrogen number
density remains almost constant from the boundary deep into the central region.
The molecular hydrogen fraction is very low in an outer shell with a thickness of
$\sim 46\:{\rm pc}$, and increases sharply from $<10^{-5}$ in the shell to
$\sim 10^{-2}$ in the center. We note that the outer shell is very thin,
compared with the depth of the center region ($\sim 1.2\:{\rm kpc}$).
The $\frac{n({\rm p})}{n({\rm HI})}$ ratio decreases monotonically with
the depth into the cloud, and the metagalactic UV radiation field does produce some
fraction of ionization in the outer shell of the cloud ($\frac{n({\rm p})}{n({\rm H})}\approx 20\%$).
Fig. 7.1 shows that the electron temperature in the center ${\rm H}_2$ core is much lower than that
in the outer shell, and has a mean value of $T_{\rm e}\approx 140\:{\rm K}$, in broad agreement
with the excitation temperature determined in Sec. 3.4.

Fig. 7.3 shows the ionization structure of carbon for our model.
The dominant ionization state is C~I and C~II.
Although the ambient UV radiation results in some fraction of C~III before it's
shielded by molecular hydrogen, the model calculation shows that the column density of
highly ionized carbon species (e.g. C~IV) is completely negligible.

Compared with observations, these results suggest that the Q1331+170 absorber has multiple
velocity components with different carbon ionization. First, the observed value
of $N_{\rm CII}/N_{\rm CI}\approx 2647$ for Q1331+170 (Liszt 2002) is much larger than
the model prediction of $\sim 3.5$ for the ${\rm H}_2$ cloud.
Second, there is strong CIV absorption associated with this damped
Ly$\alpha$ system, with $N({\rm CIV})> 10^{15}\:{\rm cm^{-2}}$ (Prochaska \& Wolfe 1999).
Therefore the bulk of the observed C~II and C~IV
are associated with other components, either in a warm/hot phase, or embedded in a
UV radiation field much stronger than that derived for the ${\rm H}_2$ absorber (see also Sec. 4.1).

Dessauges-Zavadsky {\it et al.} (2004) reported multiple velocity components
associated with the DLA of Q1331+170, and they did find
non-negligible differences between the component structure of low-ion transitions and
that of intermediate-ion transitions.
We note that the ${\rm H}_2$ cloud identified in our spectrum is different from
the five intermediate-ion velocity components by $56\:{\rm km\:s^{-1}}$, $28\:{\rm km\:s^{-1}}$,
$15\:{\rm km\:s^{-1}}$, $-15\:{\rm km\:s^{-1}}$, and $-34\:{\rm km\:s^{-1}}$
in the rest frame, respectively. On the other hand, it agrees with one low-ion component
with detected C~I absorption, with a velocity difference of only $1.6\:{\rm km\:s^{-1}}$
in the rest frame. Songaila {\it et al.}
(1994) reported two C~I absorption features associated with the same system, which
are different from the  ${\rm H}_2$ cloud by $1.4\:{\rm km\:s^{-1}}$ and
$19\:{\rm km\:s^{-1}}$ in the rest frame. The $1\:\sigma$ uncertainty is $1.2\:{\rm km\:s^{-1}}$,
caused by wavelength calibration and line centroiding (see Sec. 3.2), and therefore we conclude
that the molecular hydrogen absorption identified in our STIS spectrum is consistent with
one low-ion component reported by Dessauges-Zavadsky {\it et al.} (2004), as well as
one C~I component reported by Songaila {\it et al.} (1994).
This is as expected since ${\rm H}_2$ tends to be co-spatial with neutral carbon and/or
other low ion species. 

\subsection{Mass and dynamical state of the ${\rm H}_2$ cloud}

Based on the model shown in Fig. 7.2, we estimate the mass of the entire cloud as
$M\approx 1.36\times \frac{4}{3}\pi R^3[n({\rm HI})m_{\rm H}+n({\rm H_2})m_{\rm H_2}]\approx 6.5\times 10^7\:{\rm M_\odot}$,
where we take $R\approx 1.2\:{\rm kpc}$, $n({\rm HI})\approx 0.17\:{\rm cm^{-3}}$, 
$n({\rm H_2})\approx 0.05\:{\rm cm^{-3}}$, and the factor of $1.36$ corrects for the abundance of helium
(Dickman 1978). We refer to this mass as $M_{\rm LTE}$, since
our model is constructed based on the assumption of local thermodynamic equilibrium.
Here we have ignored the small mass fraction contributed
by H~II. We also assume spherical geometry for the absorbing region.

The masses of giant molecular clouds (GMCs) have been measured for
the Milky Way (e.g. Solomon {\it et al.} 1987, Digel {\it et al.} 1996, Heyer {\it et al.} 2001),
and nearby galaxies such as LMC (Mizuno {\it et al.} 2001a), SMC (Mizuno {\it et al.} 2001b), and M33 (Engargiola {\it et al.} 2003).
These masses are either determined based on the assumption of self-gravitational equilibrium, i.e. virial mass, $M_{\rm vir}$,
or determined by adopting a constant CO-to-${\rm H}_2$ conversion factor and then integrating the ${\rm H}_2$ column density
over the projected area of the cloud, i.e. CO luminosity mass, $M_{\rm CO}$.
In all cases, the mass spectrum of GMCs has been found to
follow approximately a power law, with the power index ranging from $\sim -1.8$ for the Milky Way (Heyer {\it et al.} 2001)
to $\sim -2.6$ for M33 (Engargiola {\it et al.} 2003). The mass of the ${\rm H}_2$ cloud associated with the
DLA towards Q1331+170 is much larger than the values determined for most GMCs in either the Milky Way or
nearby galaxies. For example, it is almost two orders of magnitude greater than the value of $7\times 10^{5}\:{\rm M_\odot}$
for the most massive GMC in M33 (Engargiola {\it et al.} 2003). Similarly,
if we adopt the GMC mass spectrum of $\frac{dM}{dN}\propto M^{-1.8}$ in the Milky Way, and
take the limiting CO luminosity of $138\:{\rm K\:km\:s^{-1}\:pc^2}$, corresponding to $566\:{\rm M_\odot}$ (Heyer {\it et al.} 2001),
we find that the fraction of GMCs with masses greater than $6.5\times 10^7\:{\rm M_\odot}$ (the value derived for the ${\rm H}_2$
cloud associated with DLA 1331+170) is only $9\times 10^{-5}$ in the Milky Way.
This may simply result from our assumption of spherical symmetry, or from
the fact that the spectral resolution of our STIS spectrum is not high enough to
resolve individual molecular clouds in velocity space along the line-of-sight.
Note that the typical velocity dispersions measured for Galactic GMCs peak at $\sigma_v\sim 1\:{\rm km\:s^{-1}}$
(Heyer {\it et al.} 2001),
much smaller than the spectral resolution of $\sim 10\:{\rm km\:s^{-1}}$ for our spectrum and the $b$-parameter
of $\sim 20\:{\rm km\:s^{-1}}$ for the H~I gas in DLA 1331+170.
The total ${\rm H}_2$ mass inside a $2\:{\rm kpc}$ galactocentric radius has been determined for a sample of
17 nearby galaxies (Paglione {\it et al.} 2001), based on the observed CO emission and a standard CO-to-${\rm H}_2$
conversion factor of $1.6\times 10^{20}\:{\rm cm^{-2}\:K^{-1}\:km^{-1}\:s}$.
These masses range from $1.6\times 10^{8}\:{\rm M_\odot}$ to $2.8\times 10^9\:{\rm M_\odot}$,
much larger than our value of $6.5\times 10^7\:{\rm M_\odot}$. This is consistent with the general deficiency of
molecular hydrogen in damped Ly$\alpha$ systems, especially when considering that the DLA towards Q1331+170
presents the largest ${\rm H}_2$ fraction in all high-z DLAs.

It is interesting to compare the mass derived above, i.e. $M_{\rm LTE}$, with the mass predicted by self-gravitational
equilibrium, i.e. $M_{\rm vir}$. We estimate the virial mass by
\begin{equation}
M_{\rm vir}\approx K\frac{\sigma_v^2 R}{G}=9.6\times 10^8\:{\rm M_\odot}.
\end{equation}
\noindent Here $\sigma_v$ is the one-dimensional velocity dispersion which we approximate as the $b$-parameter
determined in Sec. 3.2, $R\approx 1.2\:{\rm kpc}$ is the size of the cloud (see Sec. 4.2), $G$ is the gravitational
constant, and $K$ is a dimensionless factor of order unity that depends on the geometry as well as the density profile of the cloud.
We adopt $K\approx 8.7$ from Solomon {\it et al.} (1987).
Correspondingly, the gravitational parameter, $\alpha_{\rm G}=\frac{M_{\rm vir}}{M_{\rm LTE}}\approx 14$,
indicating that the ${\rm H}_2$ cloud can not be bound by self-gravity.
This is contrary to the situation for most Galactic GMCs, of which the self-gravitational equilibrium state
has long been established (e.g. Solomon {\it et al.} 1987).
More recently, Heyer {\it et al.} (2001) found that the dynamical state of Galactic GMCs varies with the cloud mass,
in that $M_{\rm CO}>10^4\:{\rm M_\odot}$ clouds are bound by self-gravity.
One possible interpretation to the large $\alpha_{\rm G}$ value
is that the ${\rm H}_2$ cloud towards Q1331+170 is in hydrostatic equilibrium,
i.e. bound by the pressure of an external medium.
Here we emphasize that the above result is subject to uncertainties due to the unknown geometry and density profile of the cloud.

\subsection{Molecular hydrogen fraction}

In Sec. 3.2, we derived a molecular hydrogen fraction of $f_{{\rm H}_2}=(5.6\pm 0.7)\%$,
for the $z=1.7765$ DLA toward Q1331+170,
which is the largest value reported so far in any redshifted damped Ly$\alpha$
absorber. The large value of $f_{{\rm H}_2}$ is consistent with the relatively high dust abundance
for this system. In Sec. 4.2, we derived the dust-to-gas ratio to be
$\tilde k\approx 0.088$. Ledoux {\it et al.} (2003)
showed that there exists a threshold at $\tilde k\approx 0.03$ above which the molecular hydrogen
fraction tends to be large ($f_{{\rm H}_2}>10^{-4}$). Moreover, the model described
in Sec. 4.2 suggests a low gas phase temperature, which is
also supported by the distribution of ${\rm H}_2$ at different $J$ states.
The low kinetic temperature is representative of a cold neutral medium (CNM) phase, in which
efficient formation of ${\rm H}_2$ molecules onto dust grains is possible. Third, 
the ambient UV radiation field of the ${\rm H}_2$ cloud is extremely low,
implying that destruction of ${\rm H}_2$ molecules by photo-dissociation is
ineffective.

\begin{figure}
\figurenum{8}
\epsscale{1.0}
\plotone{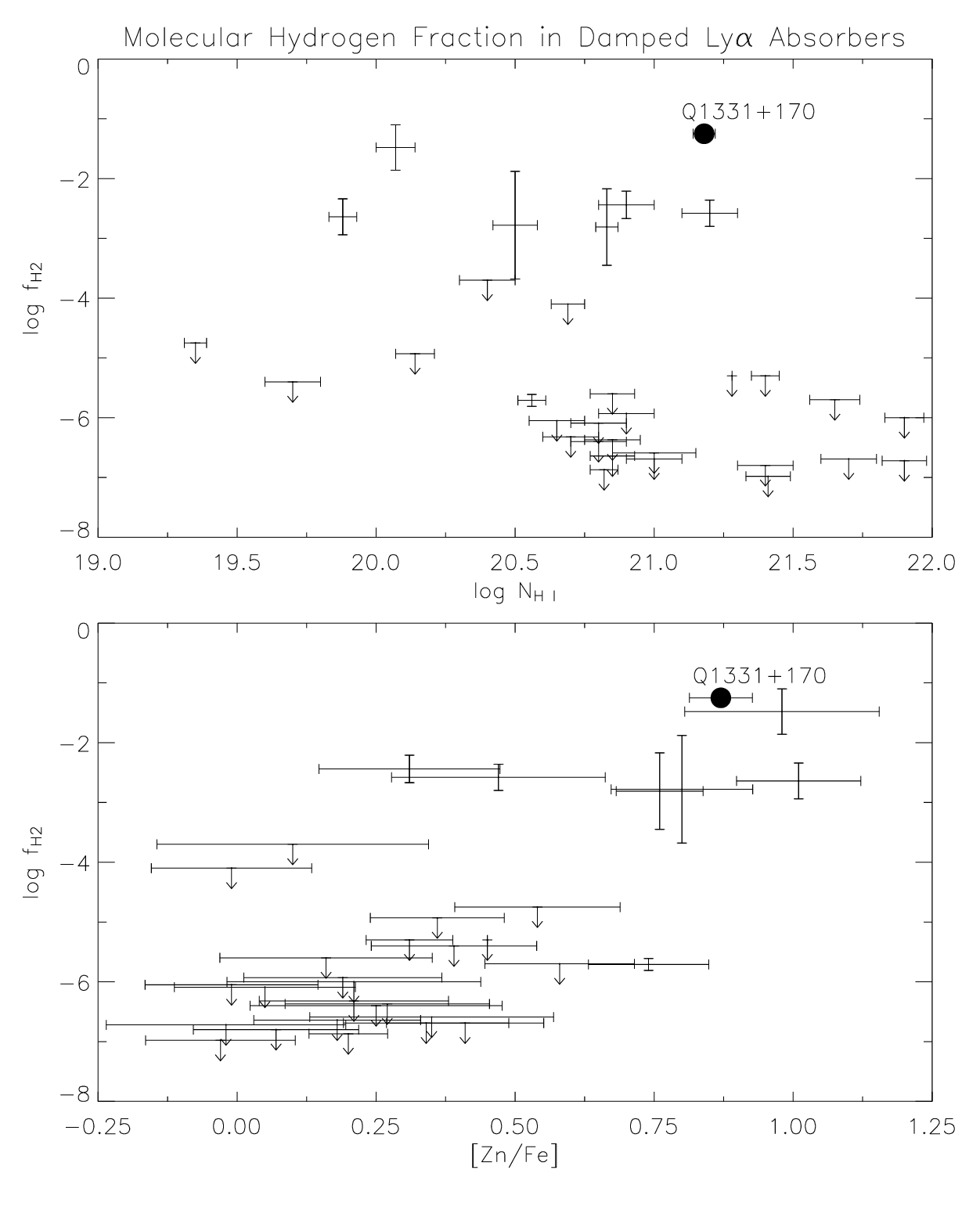}
\caption{Relation between molecular hydrogen fraction $f_{\rm H_2}$
and neutral hydrogen column density (top) and dust abundance (bottom)
for damped Ly$\alpha$ absorbers.
The upper panel shows the bi-model distribution of high-redshift DLAs,
which can be understood as a result of self-shielding of molecular hydrogen.
The lower panel shows a strong correlation between the molecular hydrogen fraction and
the dust abundance, characterized by the ratio of undepleted elements (Zn or S) to depleted elements
(Fe or Cr). Data are taken from Table 4. In both panels, the DLA
towards Q1331+170 in this paper is represented by the solid circle.}
\end{figure}

It is interesting to compare the damped Ly$\alpha$ absorber toward Q1331+170
with other high-redshift damped systems. In Table 4, we compiled data for all the 42 DLAs
which have been searched for molecular hydrogen absorption, including Q1331+170.
In the upper panel of Fig. 8, we show the distribution of damped Ly$\alpha$ absorbers with respect to neutral
hydrogen column density and molecular hydrogen fraction. The figure clearly shows the bi-model
distribution of DLAs, i.e. for a certain value of $N_{\rm HI}$,
the molecular hydrogen fraction of the absorber can be either as large as
$\sim 10^{-1}-10^{-3}$, or smaller than the detection limit ($\sim 10^{-5}-10^{-7}$).
This can be interpreted by the self-shielding effect of molecular hydrogen: once molecular
hydrogen begins to form, it shields itself from subsequent photodissociation by UV photons,
and thus the molecular hydrogen fraction increases significantly (see also Ge \& Bechtold 1999).
The lower panel of Fig. 8 shows the relation between the molecular hydrogen fraction and
the dust abundance in the same system. The dust abundance is characterized by the
column density ratio of undepleted element to depleted element. For element which is not depleted, we
take Zn as a representative, whereas when Zn is not available in the literature, we adopt
S instead. The element which is depleted to the dust grain is represented by Fe, and
when Fe abundance has not been measured, we take Cr instead.
The lower panel of Fig. 8 shows a strong correlation between the dust abundance and the
molecular hydrogen fraction, similar to previous results (e.g. Ledoux {\it et al.} 2003).
Kendall's $\tau$ test shows the correlation is present at $>99.9\%$ confidence level.
In Fig. 8, the damped Ly$\alpha$ absorber toward Q1331+170 is indicated by a solid circle.

\begin{table*}
\begin{center}
\caption{${\rm H}_2$ absorption in redshifted DLA and sub-DLA systems\label{tb4}}
\begin{tabular}{cccccccc}
\hline
Object & $z_{\rm abs}$ & ${\rm log}\:N({\rm HI})$ & ${\rm log}\:f_{\rm H_2}$ & $[\frac{\rm Zn}{\rm H}]$ $^a$ &
$[\frac{\rm Fe}{\rm H}]$ $^b$ & $\frac{E_{\rm B-V}}{0.01}$ $^c$ & Ref $^d$ \\
\hline
Q$0000-263$ & $3.390$ & $21.41\pm 0.08$ & $<-6.98$    & $-2.07\pm 0.10$ & $-2.04\pm 0.09$ & $0.00\pm 0.33$ & 2 \\
Q$0010-002$ & $2.025$ & $20.80\pm 0.10$ & $<-6.09$    & $-1.20\pm 0.12$ & $-1.25\pm 0.11$ & $0.20\pm 0.67$ & 1 \\
Q$0013-004$ & $1.973$ & $20.83\pm 0.04$ & $-2.81\pm 0.64$ & $-0.75\pm 0.06$ & $-1.51\pm 0.05$ & $4.65\pm 0.90$ & 3 \\
Q$0013-004$ & $1.968$ & $\le 19.43$     & $\ge -2.36$ & $\ge -0.73$     & $\ge -2.33$     & & 3 \\
Q$0058-292$ & $2.671$ & $21.00\pm 0.10$ & $<-6.69$    & $-1.42\pm 0.11$ & $-1.76\pm 0.10$ & $0.97\pm 0.54$ & 1 \\
Q$0100+130$ & $2.309$ & $21.40\pm 0.05$ & $<-5.30$    & $-1.62\pm 0.06$ & $-1.93\pm 0.05$ & $1.44\pm 0.45$ & 5,6 \\
Q$0102-190$ & $2.370$ & $20.85\pm 0.08$ & $<-5.60$    & $-1.73\pm 0.14$ & $-1.89\pm 0.13$ & $0.19\pm 0.24$ & 1 \\
Q$0112-306$ & $2.418$ & $20.37\pm 0.08$ & $<-5.72$    & & $-2.50\pm 0.09$ & & 1 \\
Q$0112-306$ & $2.702$ & $20.15\pm 0.07$ & $<-5.50$    & & $-0.89\pm 0.10$ & & 1 \\
Q$0112+029$ & $2.423$ & $20.70\pm 0.10$ & $<-6.32$    & $-1.14\pm 0.13$ $^*$ & $-1.35\pm 0.11$ & $0.47\pm 0.43$ & 1 \\
Q$0135-273$ & $2.800$ & $20.80\pm 0.10$ & $<-6.40$    & $-1.29\pm 0.17$ $^*$ & $-1.54\pm 0.15$ & $0.48\pm 0.49$ & 1 \\
Q$0347-383$ & $3.025$ & $20.56\pm 0.05$ & $-5.71\pm 0.10$ & $-0.98\pm 0.09$ & $-1.72\pm 0.06$ & $1.46\pm 0.41$ & 1 \\
Q$0405-443$ & $2.550$ & $21.00\pm 0.15$ & $<-6.59$    & $-1.17\pm 0.16$ & $-1.52\pm 0.15$ & $1.75\pm 1.40$ & 1 \\
Q$0405-443$ & $2.595$ & $20.90\pm 0.10$ & $-2.44\pm 0.23$ & $-1.02\pm 0.12$ & $-1.33\pm 0.11$ & $1.81\pm 1.15$ & 1 \\
Q$0405-443$ & $2.621$ & $20.25\pm 0.10$ & $<-6.15$    & & $-2.15\pm 0.10$ & & 1 \\
Q$0454+039$ & $0.860$ & $20.69\pm 0.06$ & $<-4.10$    & $-1.03\pm 0.12$ & $-1.02\pm 0.08$ & $0.00\pm 0.72$ & 6,7 \\
Q$0458-020$ & $2.040$ & $21.65\pm 0.09$ & $<-5.70$    & $-1.19\pm 0.09$ & $-1.77\pm 0.10$ & $9.95\pm 3.57$ & 5,6 \\
Q$0515-441$ & $1.151$ & $19.88\pm 0.05$ & $-2.64\pm 0.30$ & $-0.54\pm 0.05$ & $-1.55\pm 0.10$ $^{**}$ & $0.96\pm 0.28$ & 8 \\
Q$0528-250$ & $2.811$ & $21.20\pm 0.10$ & $-2.58\pm 0.22$ & $-0.78\pm 0.12$ & $-1.25\pm 0.15$ & $8.14\pm 4.14$ & 1,9,10 \\
Q$0551-366$ & $1.962$ & $20.50\pm 0.08$ & $-2.78\pm 0.90$ & $-0.15\pm 0.09$ & $-0.95\pm 0.09$ & $8.82\pm 2.73$ & 4 \\
Q$0841+129$ & $2.374$ & $20.90\pm 0.10$ & $<-5.93$    & $-1.52\pm 0.14$ & $-1.71\pm 0.11$ & $0.40\pm 0.42$ & 1,5,10 \\
Q$0841+129$ & $2.476$ & $20.65\pm 0.10$ & $<-6.05$    & $-1.52\pm 0.11$ & $-1.51\pm 0.11$ & $0.00\pm 0.23$ & 1,5,10 \\
Q$0935+417$ & $1.373$ & $20.40\pm 0.10$ & $<-3.70$    & $-0.80\pm 0.14$ & $-0.90\pm 0.20$ $^{**}$ & $0.40\pm 1.92$ & 6,11 \\
Q$1037-270$ & $2.139$ & $19.70\pm 0.10$ & $<-5.40$    & $-0.26\pm 0.11$ & $-0.65\pm 0.10$ & $0.76\pm 0.39$ & 12\\
Q$1101-264$ & $1.839$ & $19.35\pm 0.04$ & $<-4.75$    & $-0.82\pm 0.14$ $^*$ & $-1.36\pm 0.05$ & $0.08\pm 0.04$ & 1\\
Q$1117-133$ & $3.351$ & $20.85\pm 0.10$ & $<-6.37$    & $-1.28\pm 0.13$ & $-1.55\pm 0.13$ & $0.81\pm 0.62$ & 1,13\\
Q$1157+014$ & $1.944$ & $21.70\pm 0.10$ & $<-6.69$    & $-1.32\pm 0.10$ & $-1.73\pm 0.10$ & $6.86\pm 3.19$ & 1\\
Q$1223+178$ & $2.466$ & $21.40\pm 0.10$ & $<-6.80$    & $-1.63\pm 0.11$ & $-1.70\pm 0.10$ & $0.41\pm 0.89$ & 1,5\\
Q$1232+082$ & $2.338$ & $20.90\pm 0.10$ & $\ge -3.41$ & & $-1.73\pm 0.13$ & & 10,14\\
Q$1328+307$ & $0.692$ & $21.28$         & $<-5.30$    & $-1.21$         & $-1.66$         & $3.55$ & 6,15\\
Q$1331+170$ & $1.776$ & $21.18\pm 0.04$ & $-1.25\pm 0.05$ & $-1.22\pm 0.04$ & $-2.09\pm 0.04$ & $3.69\pm 0.52$ & 16\\
Q$1337+113$ & $2.508$ & $19.95\pm 0.05$ & $<-5.37$    & & $-2.05\pm 0.07$ & & 1\\
Q$1337+113$ & $2.796$ & $20.85\pm 0.08$ & $<-6.54$    & & $-2.02\pm 0.09$ & & 1\\
Q$1444+014$ & $2.087$ & $20.07\pm 0.07$ & $-1.48\pm 0.38$ & $-0.60\pm 0.15$ & $-1.58\pm 0.09$ & $1.24\pm 0.52$ & 1\\
Q$1451+123$ & $2.469$ & $20.30\pm 0.10$ & $<-4.70$    & & $-2.41\pm 0.10$ & & 1\\
Q$1451+123$ & $3.171$ & $19.90\pm 0.20$ & $<-5.80$    & & $-2.09\pm 0.24$ & & 1\\
Q$1946+766$ & $2.843$ & $20.27\pm 0.06$ & $<-5.05$    & & $-2.53\pm 0.06$ & & 5,6\\
Q$2059-360$ & $2.508$ & $20.14\pm 0.07$ & $<-4.93$    & $-1.76\pm 0.09$ $^*$ & $-2.12\pm 0.05$ & $0.02\pm 0.17$ & 1\\
Q$2059-360$ & $3.083$ & $20.85\pm 0.08$ & $<-6.64$    & $-1.65\pm 0.12$ $^*$ & $-1.83\pm 0.18$ & $0.17\pm 0.17$ & 1\\
Q$2138-444$ & $2.852$ & $20.82\pm 0.05$ & $<-6.87$    & $-1.48\pm 0.05$ & $-1.68\pm 0.05$ & $0.38\pm 0.15$ & 1\\
Q$2332-094$ & $3.057$ & $20.30\pm 0.08$ & $<-6.72$    & $-1.49\pm 0.20$ $^*$ & $-1.47\pm 0.08$ & $0.00\pm 4.32$ & 1\\
GRB $030323$ & $3.372$ & $21.90\pm 0.07$ & $<-6.00$   & $-1.26\pm 0.20$ $^*$ & $-1.47\pm 0.11$ & $5.64\pm 7.21$ & 17\\
\hline
\end{tabular}
\end{center}
$^a$ We take Zn as representative of elements undepleted by dust grains. When Zn is
not available from the literature, we take S (indicated by $^*$) instead.
$^b$ We take Fe as representative of depleted elements.
When Fe is not available from the literature, we take Cr (indicated by $^{**}$) instead.
$^c$ $E_{\rm B-V}$ values are only calculated for QSOs with foreground DLAs which have previous measurements
of both undepleted and depleted element abundances. For QSOs with more than one intervening DLAs, we only calculate the contribution
to $E_{\rm B-V}$ by individual absorbers, therefore $E_{\rm B-V}$ does not represent the total extinction of
the background QSO in some cases. These include Q0013-004, Q0405-443, Q0841+129 and Q2059-360.
$^d$ References: (1) Ledoux {\it et al.} (2003); (2) Molaro {\it et al.} (2001);
(3) Petitjean {\it et al.} (2002); (4) Ledoux {\it et al.} (2002); (5) Prochaska {\it et al.} (2001);
(6) Ge \& Bechtold (1999); (7) Pettini {\it et al.} (2000); (8) Reimers {\it et al.} (2003);
(9) Lu {\it et al.} (1996); (10) Centuri\'on {\it et al.} (2003); (11) Pettini {\it et al.} (1997);
(12) Srianand \& Petitjean (2001); (13) P\'eroux {\it et al.} (2002); (14) Srianand {\it et al.} (2000);
(15) Meyer \& York (1992); (16) Prochaska \& Wolfe (1999); (17) Vreeswijk {\it et al.} (2004).
\end{table*}

\subsection{Dust extinction}

The reddening of QSOs with intervening DLAs compared with non-DLA QSOs has been detected
and quoted as evidence for the presence of dust in DLAs (e.g. Fall \& Pei 1989,
Pei {\it et al.} 1991). More recently, Murphy \& Liske (2004) showed that there is no evidence for
the dust reddening of QSOs with DLAs, based on a much larger sample selected from the
Sloan Digital Sky Survey Data Release 2 (SDSS DR2) catalog. However, since the SDSS main QSO sample
is largely color-selected (Richards {\it et al.} 2002), the non-detection by Murphy \& Liske (2004)
may simply be a selection effect against heavily reddened objects.
In this section, we estimate the dust reddening of Q1331+170, due to the damped
Ly$\alpha$ absorber at $z\approx 1.7765$, and we also compare the reddening of QSOs for which the
intervening DLAs have different molecular hydrogen fractions.

Similar to Pei (1992), we calculate the color excess, $E_{\rm B-V}$, of Q1331+170 by
\begin{equation}
E_{\rm B-V}=1.086\frac{\kappa}{1+R_{\rm V}}(\frac{N_{\rm HI}}{10^{21}\:{\rm cm^{-2}}}),
\end{equation}
\noindent where $\kappa\approx 0.088$ for DLA 1331+170
is the dust-to-gas ratio determined from Eqn. 8,
and $R_{\rm V}$ is the ratio of total-to-selective extinction, which depends on the exact extinction curve
adopted in our calculation. If we assume the SMC extinction curve,
then $R_{\rm V}=2.93$ (Pei 1992) and $E_{\rm B-V}\approx 0.037\pm 0.005$.
We also calculate $E_{\rm B-V}$ by adopting the mean $N_{\rm HI}/E_{\rm B-V}$ values of $36.1\pm 3.3$ based on
the new determination of the SMC extinction curve (for the SMC bar sample) by Gordon {\it et al.} (2003),
where $N_{\rm HI}$ is in units of $10^{21}\:{\rm cm^{-2}}$. This gives the same value of $E_{\rm B-V}$
within $1\:\sigma$ errors. There are two reasons for
adopting the SMC extinction curve. First, the DLAs are believed to be in an early stage of chemical evolution,
which is well characterized by the SMC extinction curve (e.g. Pei 1992). Second, DLAs show no evidence
for the $2175\:{\rm \AA}$ dust feature observed in either Galactic or LMC extinction curve (e.g. Fall \& Pei 1989,
but see Malhotra 1997 which reported the detection of this feature based on a sample of Mg~II absorbers).

\begin{figure}
\figurenum{9}
\epsscale{1.0}
\plotone{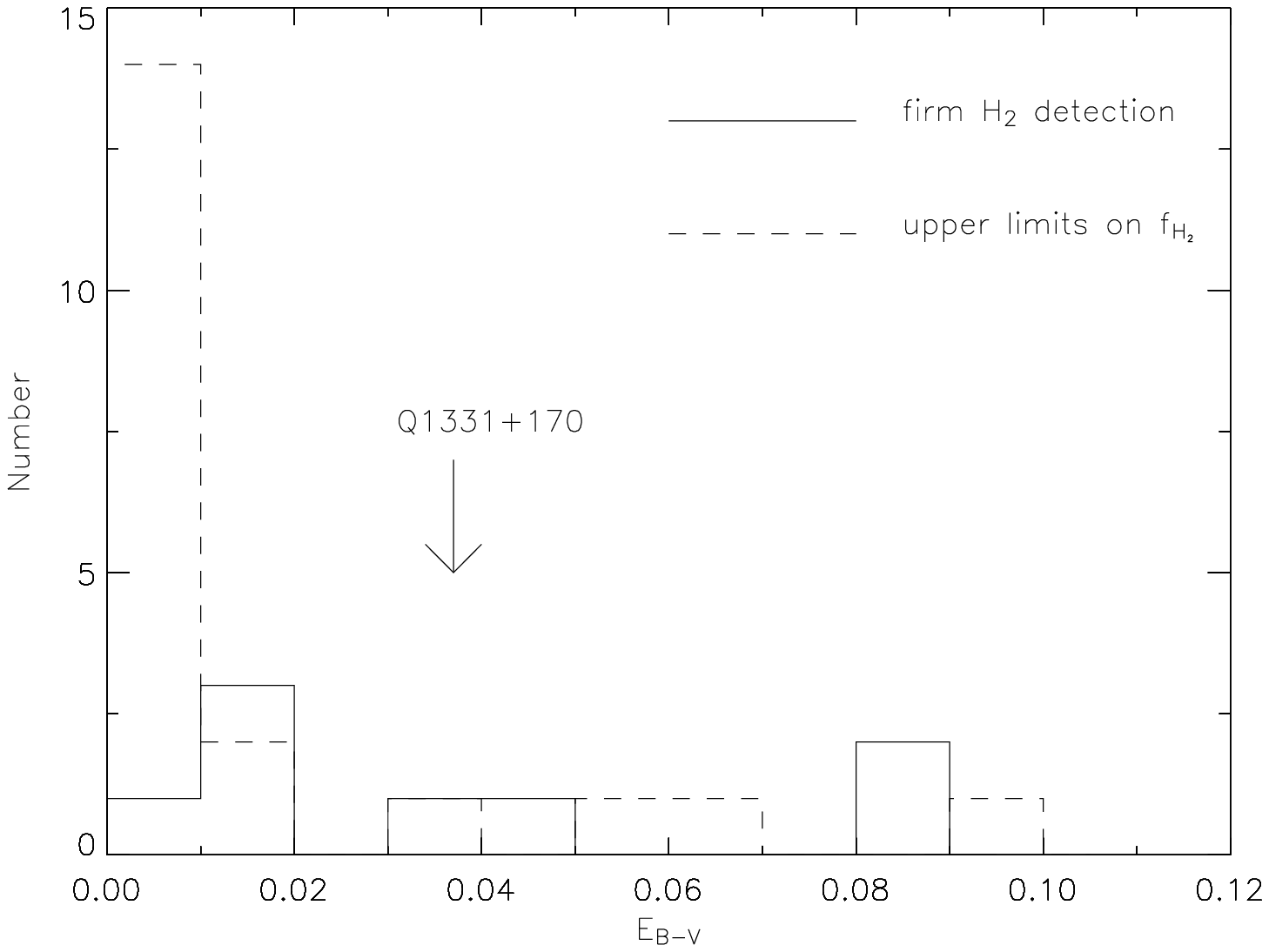}
\caption{The distribution of $E_{\rm B-V}$ for damped Ly$\alpha$ absorbers which have been
searched for molecular hydrogen absorption. The solid line represents the histogram
for the eight DLAs with firm ${\rm H}_2$ detections, whereas the dashed line represents the
histogram for other DLAs for which only upper limits of $f_{\rm H_2}$ have been put.
The two distributions are clearly different, in the sense that DLAs with ${\rm H}_2$
detections show significantly more extinction.}
\end{figure}

In column 7 of Table 4, we give $E_{\rm B-V}$ for each QSO of which the intervening DLA has previous measurements
of both depleted and undepleted metal absorption.
These values are calculated similar to the Q1331+170 case. However, we use a different
normalization when estimating the dust-to-gas ratio by Eqn. 8, depending on the types of depleted/undepleted elements
adopted in the calculation (see Table 1 of Vladilo 1998). For several QSOs with more than one intervening DLAs
(Q0013-004, Q0405-443, Q0841+129 and Q2059-360), we only list in Table 4 $E_{\rm B-V}$ contributed by individual
absorbers, which does not represent the total extinction towards the background QSO.
Fig. 9 presents the distribution of $E_{\rm B-V}$ for DLAs with firm ${\rm H_2}$ detections (solid line)
as well as the distribution for DLAs for which only upper limits of $f_{\rm H_2}$ have been put (dashed line).
The extinction for Q1331+170 is marked with an arrow.
The two distributions in Fig. 9 are different at $99.8\%$ confidence level, based on Kolmogorov-Smirnov test.
Specifically, the sub-sample of DLAs with firm ${\rm H}_2$ detections shows more extinction compared with other DLAs.
The mean color excess for absorbers with firm ${\rm H}_2$ detections is $\langle E_{\rm B-V}\rangle=0.038\pm 0.001$, significantly
greater than the value of $\langle E_{\rm B-V}\rangle=0.014\pm 0.001$ for cases in which only upper limits of $f_{\rm H_2}$
have been reported. Whereas only $12.5\%$ of DLAs with firm ${\rm H}_2$ detections have $E_{\rm B-V}$ smaller than $0.01$,
this fraction is as large as $75\%$ for other DLAs.
We have also examined the relation between $E_{\rm B-V}$ and $f_{\rm H_2}$, but no significant correlation
has been found.

The intrinsic QSO continuum in the UV/optical band is reddened by the dust extinction due to an
intervening DLA. Assuming the QSO continuum follows
$F_{\nu}\propto \nu^{-\alpha}$ or equivalently $F_{\lambda}\propto \lambda^{\alpha-2}$,
we get $\alpha\approx 4.1$ by fitting a power law to the dereddened continuum (between $2600\:\AA$
and $3000\:\AA$) of the QSO, Q1331+170, while $\alpha=4.2$ for the observed, reddened continuum adopted
in Sec. 2. For dereddening of the Galactic extinction, we use the extinction map of Schlegel {\it et al.} (1998).
For reddening of the intervening damped Ly$\alpha$ absorber at $z\approx 1.7765$, we adopt the FM
parametrization (Fitzpatrick \& Massa 1990) determined for the SMC bar sample (Gordon {\it et al.} 2003).
The observed spectrum has been shifted to the rest frame of the absorber before dereddening.
The value of $4.1$ for the dereddened spectral slope indicates an extremely red intrinsic spectrum
below the rest frame wavelength of $\sim 970\:\AA$ for Q1331+170.

\subsection{CO-to-${\rm H}_2$ ratio}

According to our model calculations in Sec. 4.2, the predicted CO column density is
$6.7\times 10^{10}\:{\rm cm^{-2}}$. This is consistent with the observed $2\:\sigma$
upper limit of $N_{\rm CO}<1.1\times 10^{13}\:{\rm cm^{-2}}$ (Levshakov {\it et al.} 1988),
determined from UV absorption lines of CO ($A^1\Pi\rightarrow X^1\Sigma^1$).
Adopting the total molecular hydrogen column density in Table 3,
we obtained the observed CO-to-${\rm H}_2$ column density ratio of
$\frac{{\rm N}_{\rm CO}}{{\rm N}_{\rm H_2}}<2.5\times 10^{-7}$. 

The upper limit of the CO-to-${\rm H}_2$ column density ratio for DLA 1331+170
is similar to the typical value of CO-to-${\rm H}_2$ measured from UV absorption of CO in Galactic diffuse
clouds, where the effects of CO self-shielding are not important (e.g. Crenny \& Federman 2004).
For dense molecular clouds in the Milky Way,
the CO/${\rm H}_2$ conversion factor is given in terms of the quantity $\frac{I_{\rm CO}}{N_{\rm H_2}}$,
where $I_{\rm CO}$ is related to the antenna temperature $T_{\rm A}$ directly measurable
from millimeter CO emission lines (e.g. Strong {\it et al.} 1988).
$\frac{I_{\rm CO}}{N_{\rm H_2}}$ usually does not provide information on the corresponding
column density ratio, since the profiles of the optically thick CO lines give the velocity
width of molecular clouds rather than the CO column densities.
The ratio of CO to ${\rm H}_2$ column densities in typical dense molecular clouds
in the Milky Way is $\sim 10^{-4}$, three orders of magnitude
higher than the upper limit determined for the DLA toward Q1331+170 (e.g. Frerking {\it et al.} 1992, Lacy {\it et al.} 1994).
This could be associated with the relatively low number density
of the ${\rm H}_2$ cloud (see Sec. 4.2), since the CO abundance is controlled by collisional reactions
involved with various species, including O, C, ${\rm H}_2$, CH and CH$_2$ (e.g. Tielens \& Hollenbach 1985).

\subsection{Measurement of CMB Temperature at $z=1.77654$}

Songaila {\it et al.} (1994) reported the detection of C~I/C~I* absorption features
associated with a velocity component at $z=1.77654$. As described in Sec. 4.2,
the redshift of this component agrees
with the redshift of the ${\rm H}_2$ absorber identified in our spectrum at $1.2\:\sigma$. 
Therefore, the C~I and ${\rm H}_2$ probably arises in the same gas cloud.
Theoretically, C~I and ${\rm H}_2$ tend to be co-spatial as well (Ge {\it et al.} 1997, 2001).
With the physical parameters determined in Sec. 4.1 and 4.2,
we can estimate the local contribution
to the C~I excitation in this system and then put constraints on the
CMB temperature at the absorber's redshift.
The equilibrium between the population
and de-population of the C~I $^3P_0$ and $^3P_1$ fine structure can be written as
\begin{equation}
N_0(B_{01}I_\nu+\Gamma_{01}+\sum_j R_{01}^j n_j)=N_1(A_{10}+B_{10}I_\nu+\Gamma_{10}+\sum_j R_{10}^j n_j),
\end{equation}
where $R_{01}$ and $R_{10}$ are collisional excitation and de-excitation rates,
with $j$ representing different collision partners (H, He, $e$, $p$, or $H_2$),
$\Gamma_{01}$ and $\Gamma_{10}$ are UV pumping
rates, $A_{10}=7.93\times 10^{-8}\:{\rm s^{-1}}$ is the probability of
the C~I fine structure transition $^3P_1\rightarrow ^3P_0$ (Bahcall \& Wolf 1968).
$B_{01}I_\nu$ and $B_{10}I_\nu$ represent excitation or de-excitation
due to the absorption of ambient microwave photons, which can be expressed as
\begin{equation}
B_{01}I_\nu=3B_{10}I_\nu=\frac{2.38\times 10^{-7}}{\exp{(23.6\:{\rm K}/T_{\rm ex})}-1}\:{\rm s^{-1}}.
\end{equation}
The collisional rates, $R_{01}^j$ and $R_{10}^j$ are taken
from Launay \& Roueff (1977), Johnson {\it et al.} (1987), Roueff \& Le Bourlot (1990),
Staemmler \& Flower (1991), and Schr\"oder {\it et al.} (1991), and are related
through the Milne relation,
\begin{equation}
R_{10}^j=\frac{1}{3}R_{01}^j\exp{(23.6\:{\rm K}/T_{\rm K})}.
\end{equation}

We take number densities of different species from the model described in Sec. 4.2.
The various collision terms are calculated to be
$R_{01}^{\rm HI}n({\rm HI})=8.6\times 10^{-10}\:{\rm s^{-1}}$,
$R_{01}^{\rm He}n({\rm He})=3.3\times 10^{-13}\:{\rm s^{-1}}$,
$R_{01}^{\rm H_2}n({\rm H}_2)=4.2\times 10^{-13}\:{\rm s^{-1}}$,
$R_{01}^{\rm p}n({\rm p})=2.3\times 10^{-12}\:{\rm s^{-1}}$, and
$R_{01}^{\rm e}n({\rm e})=1.4\times 10^{-12}\:{\rm s^{-1}}$,
assuming a kinetic temperature of $152\:{\rm K}$.
We determine the UV pumping rates for C~I by scaling from the mean Galactic values
(e.g. Ge {\it et al.} 1997), and get $\Gamma_{01}=8.7\times 10^{-13}\:{\rm s^{-1}}$
and $\Gamma_{10}=2.8\times 10^{-13}\:{\rm s^{-1}}$.
Inserting these values into Eqn. 14 and 15, and taking
$\frac{N(^3P_1)}{N(^3P_0)}=0.125\pm 0.042$ from Songaila {\it et al.} (1994),
we get $T_{\rm ex}=(7.2\pm 0.8)\:{\rm K}$ at $z=1.77654$.
Note that the uncertainty in the
calculated $T_{\rm ex}$ is associated with the errors in the measured
C~I/C~I* column densities. The excitation temperature, $T_{\rm ex}$, calculated above
is consistent with the predicted CMB temperature of $2.725\times(1+z)=7.566\:{\rm K}$
at $z=1.77654$, within $1\:\sigma$.

\section{Conclusions}

The main results of this paper are summarized as follows:

1. Using the NUV spectrum with spectral resolution of $\sim 10\:{\rm km\:s^{-1}}$
obtained with the E230M grating of HST/STIS, we detected strong
molecular hydrogen absorption associated with the $z=1.7765$ damped Ly$\alpha$ system toward
Q1331+170. The total ${\rm H}_2$ column density is $N_{\rm H_2}=(4.45\pm 0.36)\times 10^{19}\:{\rm cm^{-2}}$,
determined from simultaneous Voigt profile fitting of 26 ${\rm H}_2$ lines.

2. The molecular hydrogen fraction was determined to be
$f_{\rm H_2}=\frac{2N_{\rm H_2}}{N_{\rm HI}+2N_{\rm H_2}}=(5.6\pm 0.7)\%$,
which is the largest value reported so far
in any redshifted damped Ly$\alpha$ system. This is a combined effect of a relatively high
dust-to-gas ratio, a low gas temperature, and an extremely low ambient UV radiation field.

3. We detect rotationally excited transitions of ${\rm H}_2$, with $J=0 - 4$. The relative column
densities of the $J=0,1,2,3,4$ states can be fit with a single temperature of $T_{\rm ex}=152\pm 10\:{\rm K}$.
This suggests that the derived excitation
temperature represents the kinetic temperature of the gas, and that photo-excitation is negligible.

4. The photo-absorption rate was estimated to be
$R_{\rm abs}=(7.6\pm 2.4)\times 10^{-13}\:{\rm s^{-1}}$, corresponding to an
ambient UV radiation field of $J_{912{\rm \AA}}\approx 7\times 10^{-23}\:{\rm ergs\:s^{-1}\:cm^{-2}\:Hz^{-1}\:Sr^{-1}}$.
This is comparable with the metagalactic UV background intensity at the same redshift,
and implies an extremely low star-formation rate in the DLA environment. 

5. We constructed a simple model to describe the structure of the ${\rm H}_2$ absorber,
with a best-fit total hydrogen number density of $n_{\rm H}\approx 0.2\:{\rm cm^{-3}}$
and an electron temperature of $T_{\rm e}\approx 140\:{\rm K}$. The total mass of the
model cloud is  $\sim 6.5\times 10^7\:{\rm M_\odot}$, greater than masses of most GMCs in the Milky
Way and nearby galaxies. We also find that this mass is considerably smaller than the virial mass of the cloud,
indicating that the ${\rm H}_2$ cloud is not in a state of self-gravitational equilibrium.
Our model is consistent with the population of ${\rm H}_2$
at different $J$ states, as well as the observational upper limit put on the
CO column density. The observed CO-to-${\rm H}_2$ column density ratio is
$\frac{{\rm N}_{\rm CO}}{{\rm N}_{\rm H_2}}<2.5\times 10^{-7}$, characteristic of
the typical value measured for diffuse molecular clouds in the Galactic ISM.

6. Adopting the SMC-like extinction curve, we calculate the extinction of Q1331+170 to be $E_{\rm B-V}=0.037\pm 0.005$,
due to the intervening damped Ly$\alpha$ absorber at $z\approx 1.7765$.
We also find that the extinction of QSOs by foreground DLAs with firm ${\rm H}_2$ detections is
considerably greater than those for which only upper limits of $f_{\rm H_2}$ have been put.

7. The redshift of the ${\rm H}_2$ absorber identified in our spectrum is consistent with a C~I/C~I*
absorber reported by Songaila {\it et al.} (1994). We calculated the local contribution
to C~I excitation, including UV pumping and collisions. The residual excitation temperature
was determined to be $T_{\rm ex}=(7.2\pm 0.8)\:{\rm K}$, consistent with
the predicted $T_{\rm CMB}$ of $7.566\:{\rm K}$ at $z=1.77654$.

\acknowledgments

We are grateful to R.I. Thompson, D.J. Eisenstein, X. Fan, D. Welty, J. Black, 
R. Dav\'e, J. Bieging and the anonymous referee for their suggestions, which have greatly improved this work.
We also thank R.F. Carswell and G. Ferland for making the Voigt profile fitting program VPFIT and
the photoionization code CLOUDY publicly available. This work is based on observations made with
the NASA/ESA {\it Hubble Space Telescope}, obtained at the {\it Space Telescope Science Institute},
which is operated by the {\it Association of Universities for Research in Astronomy},
Inc., under NASA contract NAS 5-26555. These observations
were supported by NASA through Grants HST-GO-09172.01 and HST-GO-07271.01 from
the {\it Space Telescope Science Institute}.

\end{document}